\documentclass[manuscript]{aastex}
\usepackage{amsmath,amssymb}
\usepackage{epstopdf}
\usepackage{graphicx}
\usepackage{xcolor}

\usepackage[titletoc,title]{appendix}
\title{Characterization of Turbulent Fluctuations in the Sub-Alfv\'enic Solar Wind}

\newcommand{\myemail}

\usepackage[inline]{enumitem}
\usepackage{multirow}
\usepackage{lscape}

\shorttitle{}
\shortauthors{Zank et al.}

\begin{document}

\begin{abstract}
Parker Solar Probe (PSP) observed sub-Alfv\'enic solar wind intervals during encounters 8 - 14, and  low-frequency magnetohydrodynamic turbulence in these regions may differ from that in super-Alfv\'enic  wind. We apply a new mode-decomposition analysis \citep{Zank_etal_2023} to the sub-Alfv\'enic  flow observed by PSP on 2021 April 28, identifying and characterizing entropy, magnetic islands, forward and backward Alfv\'en waves, including weakly/non-propagating Alfv\'en vortices, forward and backward fast and slow magnetosonic modes. Density fluctuations are primarily and almost equally  entropy and backward propagating slow magnetosonic modes. The mode-decomposition provides phase information (frequency and wavenumber $k$) for each mode. Entropy-density fluctuations have a wavenumber anisotropy $k_{\parallel} \gg k_{\perp}$ whereas slow mode density fluctuations have $k_{\perp} > k_{\parallel}$. Magnetic field fluctuations are primarily magnetic island modes ($\delta B^i$) with an $O(1)$ smaller contribution from uni-directionally propagating  Alfv\'en waves ($\delta B^{A+}$) giving a variance anisotropy of $\langle {\delta B^i}^2 \rangle / \langle {\delta B^A}^2 \rangle = 4.1$. Incompressible magnetic fluctuations dominate compressible contributions from fast and slow magnetosonic modes. The magnetic island spectrum is Kolmogorov-like $k_{\perp}^{-1.6}$ in perpendicular wavenumber and the uni-directional Alfv\'en wave spectra are $k_{\parallel}^{-1.6}$ and $k_{\perp}^{-1.5}$.  Fast magnetosonic modes propagate at essentially the Alfv\'en speed with anti-correlated transverse velocity and magnetic field fluctuations and are almost exclusively magnetic due to $\beta_p  \ll 1$. Transverse velocity fluctuations are the dominant velocity component in fast magnetosonic modes and longitudinal fluctuations dominate in slow modes.  Mode-decomposition is an effective tool in identifying the basic building blocks of MHD turbulence and provides detailed phase information about each of the modes.
\end{abstract}
	
	\author{G.P.\ Zank$^{1,2}$, L.-L. Zhao$^{1,2}$, L.\ Adhikari$^{1,2}$, D.\ Telloni$^3$, Prashant Baruwal$^{1,2}$, Prashrit Baruwal$^{1,2}$, Xingyu Zhu$^1$, M. Nakanotani$^1$, A.\ Pit\u{n}a$^4$, J. C. Kasper$^5$, S. D. Bale$^6$  }
	
	\altaffiltext{1}{Center for Space Plasma and Aeronomic Research (CSPAR), University of Alabama in Huntsville, Huntsville, AL 35899, USA }
	
	\altaffiltext{2}{Department of Space Science, University of Alabama in Huntsville, Huntsville, AL 35899, USA}	
	
	\altaffiltext{3}{National Institute for Astrophysics, Astrophysical Observatory of Torino, Via Osservatorio 20, I-10025 Pino Torinese, Italy}
	\altaffiltext{4}{Department of Surface and Plasma Science, Faculty of Mathematics and Physics, Charles University, Prague, Czechia}	
	\altaffiltext{5}{BWX Technologies, Inc., Washington DC 20002, USA and Department of Climate and Space Sciences and Engineering, University of
		Michigan, Ann Arbor, MI 48109, USA}
	\altaffiltext{6}{Physics Department, University of California, Berkeley, CA 94720-7300, USA}

\section{Introduction} \label{sec:1} 

The first entry of the NASA \textit{Parker Solar Probe} (PSP) spacecraft into a sub-Alfv\'enic solar wind flow lasted for 5 hours on 2021 April 28, and has been studied extensively \citep{Kasper_etal_2021, Zank_etal_2022, Zhao_etal_2022, Bandyopadhyay_etal_2022, Alberti_etal_2022, Zhang_etal_2022, Liu_etal_2023, Jiao_etal_2024}.  Unlike the super-Alfv\'enic solar wind, the turbulence properties of this new solar wind environment are connected magnetically to the surface of the Sun. Sub-Alfv\'enic solar wind intervals have now been observed since encounter 8 through to encounter 14. \cite{Jiao_etal_2024} have traced magnetic fields to locate the source of the sub-Alfv\'enic intervals or streams, finding that the sources are either the boundaries inside coronal holes or small regions of open magnetic field. They find that the location of the Alfv\'en surface for these flows varies between 15 and 24 solar radii $R_{\odot}$, which is larger than the canonical value of 11 - 12 $R_{\odot}$ assumed typically for fast solar wind originating from an open coronal hole.  \cite{Jiao_etal_2024} find that the sub-Alfv\'enic intervals so far observed all exhibit similar properties and origins. 

While potentially different from the physics governing large open coronal holes or closed magnetic field regions, it is nonetheless instructive to better understand the nature of turbulence and the potential origin of super-Alfv\'enic flows in the sub-Alfv\'enic regions that PSP has so far discovered. 
The focus on the properties of turbulence in this unexplored region of the solar wind reflects the idea that the dissipation of low-frequency magnetohydrodynamic (MHD) turbulent fluctuations provides the distributed heating source for the solar corona that results in the driving of the solar wind. Two basic coronal turbulence models have been advanced, one treating the turbulence as predominantly slab \citep{matthaeus_etal_1999_coronalheating} and the other as primarily non-propagating 2D nonlinear structures such as small scale magnetic flux ropes and Alfv\'en vortices plus a minority slab component \citep{Zank_etal_2018}, both of which have been reviewed in \cite{Zank_etal_2021}. Both models have since been refined and extended by, e.g., \cite{oughton_etal_2001_reducedmhdmodel, dmitruk_etal_2001, Dmitruk_etal_2002, Suzuki_Inutsuka_2005, Cranmer_etal_2007, Cranmer_van_Ballegooijen_2012, Cranmer_etal_2013, Wang_etal_2009, Chandran_Hollweg_2009, Verdini_etal_2010, Matsumoto_Shibata_2010, Chandran_etal_2011, usmanov_etal_2011_swmodturbtransheat, Lionello_etal_2014, Usmanov_etal_2014, Woolsey_Cranmer_2014, Shoda_etal_2018, Chandran_Perez_2019, Chandran_2021} for Alfv\'en wave or slab turbulence heating and driving, and by \cite{Zank_etal_2021, Adhikari_etal_2020a, Adhikari_etal_2022c, Telloni_etal_2022a, Telloni_etal_2022b, Telloni_etal_2023a} for 2D nonlinear structures. PSP observations  \citep{Bale_etal_2023, Raouafi_etal_2023}, Solar Dynamics Observatory (SDO) spacecraft observations  \citep{Uritsky_etal_2023}, and theory \citep{Zank_etal_2018, Priest_etal_2018, Pontin_etal_2024} have identified small- and multi-scale magnetic reconnection low in the corona as a possible mechanism for solar coronal heating. Such small- and multi-scale reconnection can be expected to generate slab \citep{matthaeus_etal_1999_coronalheating}, 2D \citep{Zank_etal_2018}, or an admixture of magnetohydrodynamic (MHD) turbulence low in the corona. Exactly which form is dominant, if any, is not known. 

Identifying the underlying character of the turbulent fluctuations in the coronal flow is key to distinguishing between the two competing turbulence models of solar coronal heating. The turbulence models are based typically on either fully incompressible 3D MHD regardless of plasma beta or nearly incompressible MHD in which the plasma beta $\beta_p$ ($= P/(B^2/2\mu_0)$ where $P$ is the plasma pressure, $B = |{\bf B}|$, ${\bf B}$ the magnetic field, and $\mu_0$ the magnetic permeability) distinguishes the leading-order incompressible description (2D incompressible MHD for $\beta_p \ll 1$ or $O(1)$ or 3D incompressible MHD for $\beta_p \gg 1$) \citep{Zank_Matthaeus_1992b,  zank_matthaeus_1993_incompress, Zank_etal_2017a}. The associated spectral anisotropy of incompressible 3D MHD is typically expressed via the \cite{Goldreich_Sridhar_1995} critical balance theory in which it is hypothesized that the nonlinear and Alfv\'enic timescales are balanced. This yields e.g., a reduced 1D wavenumber (${\bf k}$) spectrum for Els\"asser fluctuations of the form $k_{\perp}^{-5/3}$ and $k_{\parallel}^{-2}$, where $k_{\perp} = |{\bf k}_{\perp}|$ and $k_{\parallel}$ are wavenumbers perpendicular and parallel to the mean magnetic field ${\bf B}_0$ provided the normalized cross helicity $\sigma_c = 0$ (where $\sigma_c = {\bf u} \cdot {\bf b} /\sqrt{\mu_0 \rho_0} / \left(u^2 + b^2 /(\mu_0 \rho_0) \right)$ and ${\bf u}$ is the fluctuating velocity and ${\bf b} /\sqrt{\mu_0 \rho_0} $ the fluctuating Alfv\'en velocity, and $\rho_0$ the mean fluid density). There is mounting evidence  \citep{Telloni_etal_2019, Wang_etal_2015, Zhao_etal_2021a, Zank_etal_2022} that highly field-aligned flows with $|\sigma_c| \simeq 1$, i.e., populated by uni-directionally propagating Alfv\'en waves have reduced 1D spectra of the form $k_{\parallel}^{-5/3}$ to $k_{\parallel}^{-3/2}$, in contrast to the predictions of critical balance. However, based on the spectral theory developed in \cite{Zank_etal_2020} for the 2D + slab superposition model with a dominant 2D component \citep{matthaeus_etal_1990_quasi2dfluct, Bieber_etal_1994, Bieber_etal_1996JGR, Sauer_Bieber_1999, Forman_etal_2011} such spectra can be explained \citep{Zank_etal_2022, Zhao_etal_2022a, Zhao_etal_2022b, Zhao_etal_2022} as a consequence of sweeping \citep{Zhao_etal_2023a} of the slab turbulence by the dominant 2D modes, which was described as ``scattering'' in \cite{Zank_etal_2020}.

To distinguish between these two descriptions of low-frequency turbulence, one would ideally like to separate Alfv\'en waves from advected structures. In a similar vein, one would like to distinguish between density fluctuations \citep[e.g.,][]{Kontar_etal_2023} generated by compressible fast and slow magnetosonic waves and advected density fluctuations (entropy modes). \cite{Zank_etal_2023} revisited and developed a very general mode-decomposition method that identifies wave modes and 
advected structures such as magnetic islands or entropy modes while evaluating the corresponding phase information. By utilizing the newly developed mode-decomposition technique, we re-analyze the first sub-Alfv\'enic solar wind interval observed by PSP. In so doing, we 1) identify all the possible low-frequency MHD modes, including advected modes, that are present in the observed 5-hour plasma parcel, and 2) derive the spectral characteristics of all the identified modes. Of particular note is that, because the analysis resides within a linear framework, we can exploit the corresponding dispersion relations for each mode to relate frequency to wavenumber and thereby present wavenumber spectra (reduced power spectral densities (PSDs) as functions of $k$, $k_{\perp}$, and $k_{\parallel}$) without invoking Taylor's hypothesis for the propagating or wave modes. From this analysis, a clear characterization of low-frequency inertial range MHD fluctuations is obtained in the sub-Alfv\'enic solar wind, from which the relative contribution of the various components can be extracted. 

It should be noted that a ``classical'' mode-decomposition analysis \citep{Glassmeier_etal_1995} was applied to various plasma intervals observed during the first PSP encounter by \cite{Chaston_etal_2020} and \cite{Zhao_sq_etal_2021}. Unlike the method introduced by \cite{Zank_etal_2023}, the classical method projects the observed fluctuations onto a subspace of possible MHD modes comprising only three possible modes, the planar Alfv\'en, and fast and slow ms modes, and all three modes share the same wave vector \citep{Zhao_sq_etal_2021}. The classical approach does not include advected modes (entropy modes, magnetic islands, Alfv\'en vortices) nor the full spherical Alfv\'en mode, all of which are incorporated in the analysis presented in \cite{Zank_etal_2023}, together with the fast and slow ms modes. The \cite{Zank_etal_2023} analysis yields  the phase information for each possible mode, therefore relating fluctuations observed at a particular frequency to the wavenumber of the specific mode. Because the \cite{Chaston_etal_2020} and \cite{Zhao_sq_etal_2021} analyses project the observed fluctuations onto a subspace of the full MHD mode space, one cannot meaningfully compare their conclusions, limited as they are by the assumptions underlying the analysis, to the detailed results presented here. A mode-decomposition analysis of the super-Alfv\'enic young solar wind is therefore underway using the \cite{Zank_etal_2023} approach to effect such a comparison. 

We emphasize that the mode-decomposition analysis is not a theory of turbulence in any sense but is rather a tool to identify small amplitude MHD fluctuations in a particular plasma parcel, from which we can identify in part the building blocks of a turbulent fluid. 

In the following section, we review briefly the new mode-decomposition technique, focusing particularly on how to apply the method to sub-Alfv\'enic flows. Thereafter, in \S \ref{sec:3}, we identify the MHD modes present in the first 5-hour sub-Alfv\'enic solar wind flow and present their detailed spectral characteristics. The implications of our results for turbulence models thought to be responsible for heating the solar corona are discussed in \S \ref{sec:4}. The algorithm and formulae used in the mode-decomposition analysis are listed in the Appendix, correcting some typos that were present in the original paper \citep{Zank_etal_2023}. 

\section{MHD Mode-decomposition} \label{sec:2} 

Here we expand the mode-decomposition technique for MHD introduced in \cite{Zank_etal_2023} to sub-Alfv\'enic flows. The primary difference between this and the prior analysis is in the identification of the time intervals to ensure that the linear analysis remains valid for the time over which the decomposition is evaluated. Specifically, in analyzing the superposition of the various fluctuating modes, the normal modes expansion of the measured plasma or magnetic field fluctuation about a ``decoherence'' time $\Delta t$ must ensure that the decomposition extracts a coherent superposition of the constituent modes. Here we consider a plasma parcel in a modestly super- or sub-Alfv\'enic flowing medium such as the solar wind, restricting our attention to relatively quiescent flows that do not contain shocks, large-scale heliospheric current sheet crossings, and other unspecified large-amplitude/non-linear events. 
In analyzing a fluid parcel of length $\ell$ say, loosely defined to be along the radial direction, which in the coronal case is roughly parallel to the mean magnetic field (in principle, we need to consider the radial, tangential, and normal directions more generally  but the argument carries over easily), we need to consider three timescales that help determine $\Delta t$, these being 1) the characteristic  timescale $\tau_{prop}$ for fluctuations to propagate out of the plasma parcel; 2) the nonlinear timescale $\tau_{nl}$, and 3) the Alfv\'en timescale $\tau_A$.  

For the mode-decomposition to apply, we need to ensure that the waves do not propagate outside the plasma parcel during the time $\Delta t$ of the analysis (point 1 above), and secondly (points 2 and 3) that the fluctuations in the parcel do not undergo nonlinear interactions and thus violate linearity. 
This imposes the following conditions on the choice of $\Delta t$. To address 1), the phase speed of a propagating fluctuation in the fluid frame is $V_{ph} = \omega /k$ where $\omega$ is the wave frequency (advected fluctuations do not need to be considered obviously). Hence, we have $\tau_{prop}  \sim \ell /(\omega /k)$, which yields the requirement that the time intervals $\Delta t \ll \tau_{prop} \sim \ell /(\omega /k)$. Since $\ell = |{\bf U}| T$, where $T$ is the duration of the plasma parcel as it is advected past the spacecraft at a velocity ${\bf U}$ (i.e., the relative speed of the plasma flow and the spacecraft), we therefore require $\Delta t \ll |{\bf U}| T / (\omega /k)$. For highly super-Alfv\'enic and supersonic flows, $|{\bf U}| \sim U_0 \gg V_{A0}$ ($U_0$ and $V_{A0}$ are the mean flow and Alfv\'en speed respectively), we therefore have the condition that $\Delta t/T \ll U_0 /(\omega /k)$ which is $\gg 1$ for a supersonic and super-Alfv\'enic flow. However, for an Alfv\'enic flow $U_0 \sim V_{A0}$ in the solar corona where $\beta_p \ll 1$ typically (implying that the Alfv\'en and fast magnetosonic speeds are similar since $V_p^{f\pm} \simeq V_{A0} \left[ 1 + (\gamma /4) \beta_p \sin^2 \theta^{f\pm} \right]$ when $\beta_p \ll 1$, and is much large than the slow magnetosonic speed), being about $0.1$ for this particular interval, we have instead that $\Delta t/T \ll (U_0 + V_{A0}) /(\omega /k) \sim 2$ since smaller scale fluctuations will be swept by longer wavelength fluctuations, i.e., $|{\bf U}| \simeq U_0 + V_{A0}$. Hence, for modestly super- or sub-Alfv\'enic flows, we require $\Delta t /T \ll O(1)$. Thus, for a plasma parcel of size $\ell \simeq U_0 T$, the full parcel for an Alfv\'enic flow must be broken into subintervals to ensure $\Delta t \ll T$ (see Figure 12 in \cite{Zank_etal_2023}) and the wave mode-decomposition  applied to each subinterval as described in \cite{Zank_etal_2023}, Figure 12. Secondly, the nonlinear timescale is expressed as $\tau_{nl}^{-1} \sim \langle z^2 \rangle^{1/2} k$, where $z^2$ is the fluctuating Els\'asser energy. We require that $\Delta t \ll  \tau_{nl} \sim \langle z^2 \rangle^{-1/2} k^{-1} \sim  \ell /\langle z^2 \rangle^{1/2}$. This yields the condition $\Delta t \ll (U_0 + V_{A0}) T /\langle u^2 \rangle^{1/2}$. For $U_0 \gg V_{A0}$, $\Delta t/T \ll U_0/\langle u^2 \rangle^{1/2}$ ($\gg 1$), and for $U_0 \sim V_{A0}$,  $\Delta t /T \ll 2V_{A0} /\langle u^2 \rangle^{1/2} \sim 2U_{0} /\langle u^2 \rangle^{1/2}$ ($\gg 1$). 

Finally, the Alfv\'en timescale for nonlinear interactions can be expressed as $\tau_A^{-1} = V_{A0} k (1 - {\sigma_c^A}^2 )$ \citep{Zank_etal_2020}, where $\sigma_c^A$ is the cross helicity of Alfv\'enic fluctuations. For approximately equally counter-propagating Alfv\'en modes, $\sigma_c^A \simeq 0$ and the Alfv\'en timescale is the familiar 
$\tau_A^{-1} \sim V_{A0} k$. In this case, we require that $\Delta t \ll 1/(V_{A0} k)< \ell /V_{A0} \sim (U_0 + V_{A0}) T /V_{A0}$. As before, for super-Alfv\'enic flows $U_0 \gg V_{A0}$ implies $\Delta t/T \ll U_0 /V_{A0}$ ($\gg 1$) whereas for $U_0 \sim V_{A0}$ we have $\Delta t/T \ll O(1)$, and in the latter case, the analysis needs to be done on a series of subintervals. By contrast, for unidirectionally propagating Alfv\'en modes, $|\sigma_c^A| \simeq 1$ and hence $\tau_A^{-1} = 0$, i.e., nonlinear interactions do not occur (other than via ``sweeping'' or ``scattering'' - see \cite{Zank_etal_2017a, Zank_etal_2020} which we do not address here), in which case only the timescale $\tau_{prop}$ is  relevant. 

$T$ is fixed and corresponds to the period of the observation, which in this case  was 5 hours for the  sub-Alfv\'enic wind. During those 5 hours, there were only relatively small changes in the basic background plasma and magnetic field parameters so one can reasonably consider the whole interval. Had there been major changes, such as a shock wave or a heliospheric current sheet crossing, the corresponding ``smooth and relatively unchanging'' period of the plasma observed would have defined $T$. As described above, it is necessary and sufficient to choose $\Delta t$ to satisfy $\Delta t /T \ll 1$, so since $T = 5$ hours, choosing $\Delta t = 30$ minutes implies $\Delta t/T = 0.1$, which should be a sufficiently small interval while retaining a reasonable level of statistical accuracy. 

We consider the first sub-Alfvenic interval \citep{Kasper_etal_2021}  measured by the FIELDS \citep{Bale_etal_2016} and SWEAP \citep{Kasper_etal_2016} instruments on the NASA Parker Solar Probe spacecraft \cite{Fox_etal_2016}, observed between 09:30 and 14:40 UT on 2021 April 28 at $\sim 0.1$ au. The basic plasma parameters for this period are depicted in Figure 1 of \cite{Zank_etal_2022} between the dashed lines. During this time, $M_A \equiv U/V_A \leq 1$ and $\beta_p \sim 10^{-1}$. Hence, the fast magnetosonic speed is only a little greater than $V_A$. Since both the propagation and Alfv\'en timescales impose the decoherence time condition $\Delta t /T \ll O(1)$, we subdivide the 5-hour interval into 10 30-minute intervals, calculate the relevant means of the plasma and magnetic field variables for each, and then perform the corresponding mode-decomposition analysis on each of the subintervals. As in \cite{Zank_etal_2023} (see their Figure 1), the coordinate system in each subinterval uses the mean magnetic field ${\bf B}_0 = B_0 \hat{\bf z}$ to define the $\hat{\bf z}$ direction, $\hat{\bf x}$ and $\hat{\bf y}$ complete the triad, and the mean flow velocity ${\bf U}_0$ is, without loss of generality,  rotated into the $(x,z)$-plane. For completeness and to correct some typos in \cite{Zank_etal_2023}, the mode-decomposition algorithm is listed in the Appendix. Based on the re-constructed subinterval mode-decomposition analysis, we perform a spectral analysis of the individual low-frequency MHD modes that are present in the 5-hour sub-Alfv\'enic interval.

\section{Decomposition Results} \label{sec:3}

As described in \cite{Zank_etal_2023}, after subdividing the 5-hour interval into 10 30-minute intervals, we compute mean values within each interval for the plasma and magnetic field variables, and then apply the mode-decomposition algorithm to identify the following MHD linear/small-amplitude modes;  entropy modes, forward ($+$) and backward ($-$) fast ($f$) and slow ($s$) magnetosonic (ms) modes, magnetic island ($i$) or flux rope modes, and forward ($+$) and backward ($-$) Alfv\'en ($A$) modes within each interval. The mode-decomposition evaluates both the amplitude and the phase $(\omega, {\bf k})$ ($\omega$ the frequency and ${\bf k}$ the wavenumber) of each MHD mode. 

\begin{figure}
	\includegraphics[width= 1.0\textwidth]{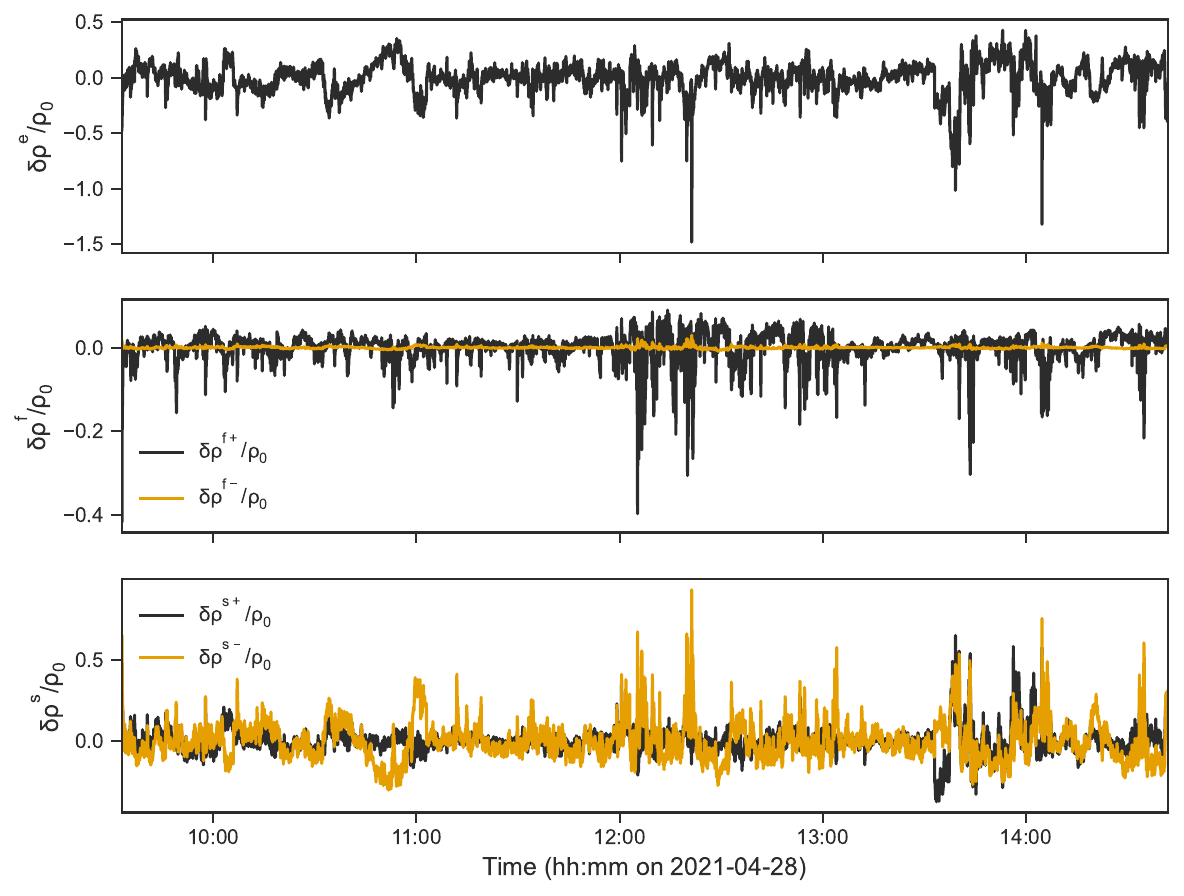}
	\caption{\small Time series of the normalized density fluctuations over the composite 10 30-minute subintervals of the 5-hour sub-Alfv\'enic interval.  {\bf Top panel:} Normalized density fluctuations for the entropy modes. {\bf Middle panel:} Normalized density fluctuations for the forward (black curve) and backward (orange curve) fast magnetosonic modes. {\bf Bottom panel:} Normalized density fluctuations for the forward (black curve) and backward (orange curve) slow magnetosonic modes. } \label{fig:1}
\end{figure}

\begin{figure}
	$$\includegraphics[width= 0.3\textwidth]{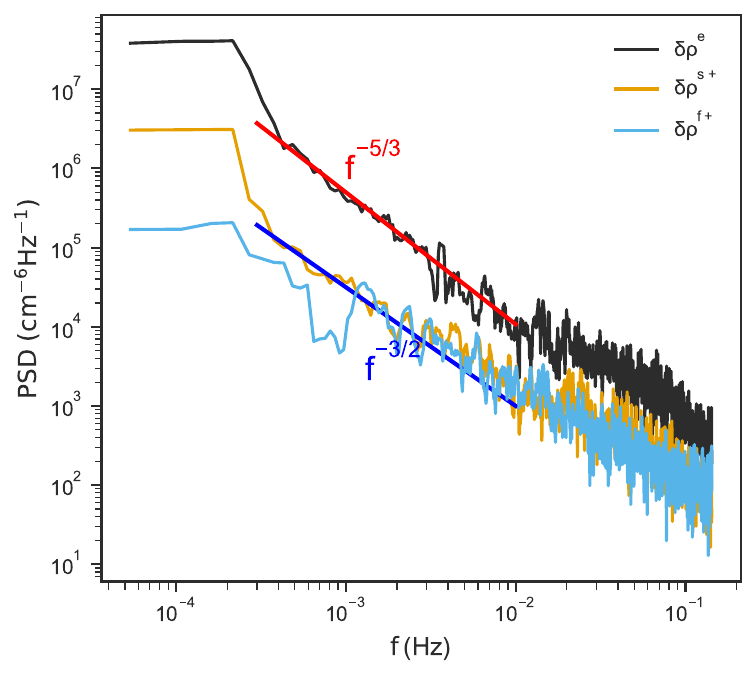}
	\includegraphics[width= 0.3\textwidth]{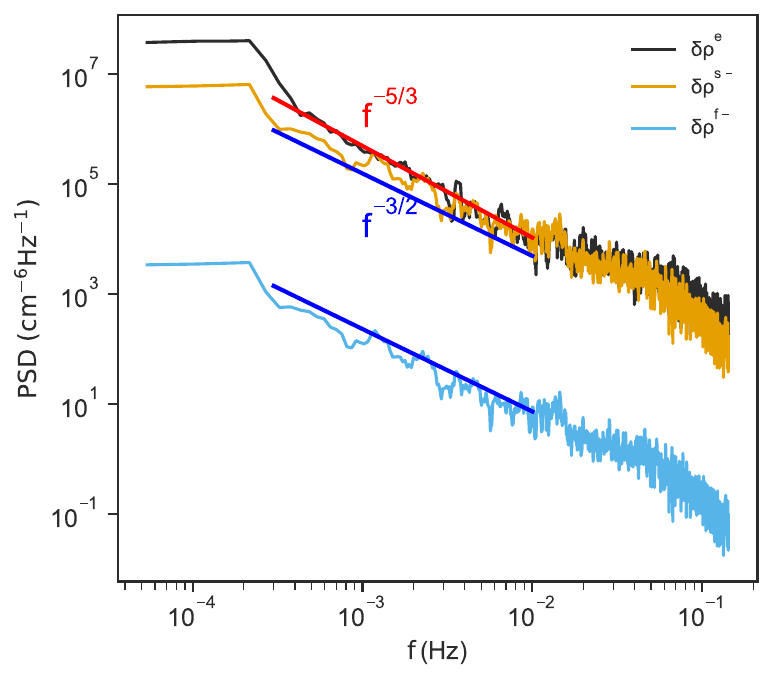}$$
	$$\includegraphics[width= 0.3\textwidth]{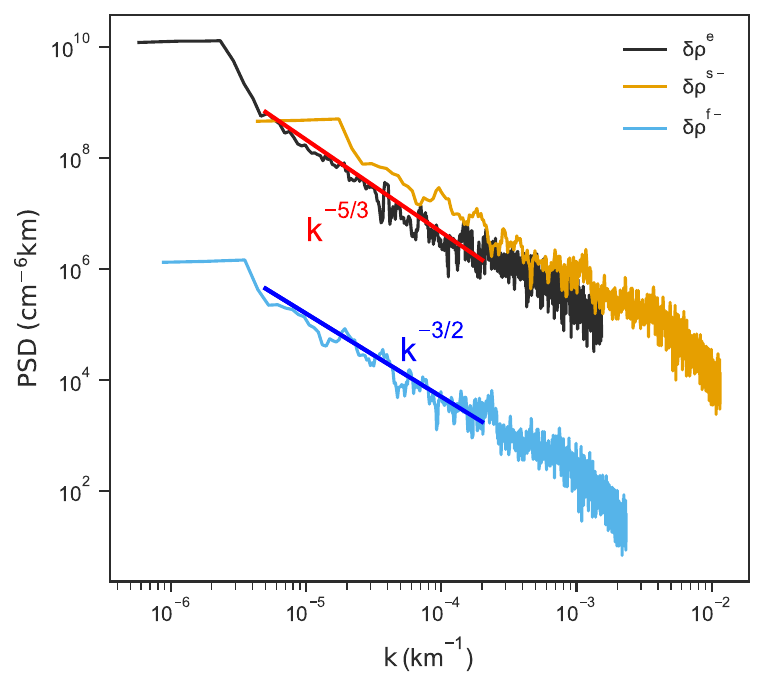}
	\includegraphics[width= 0.3\textwidth]{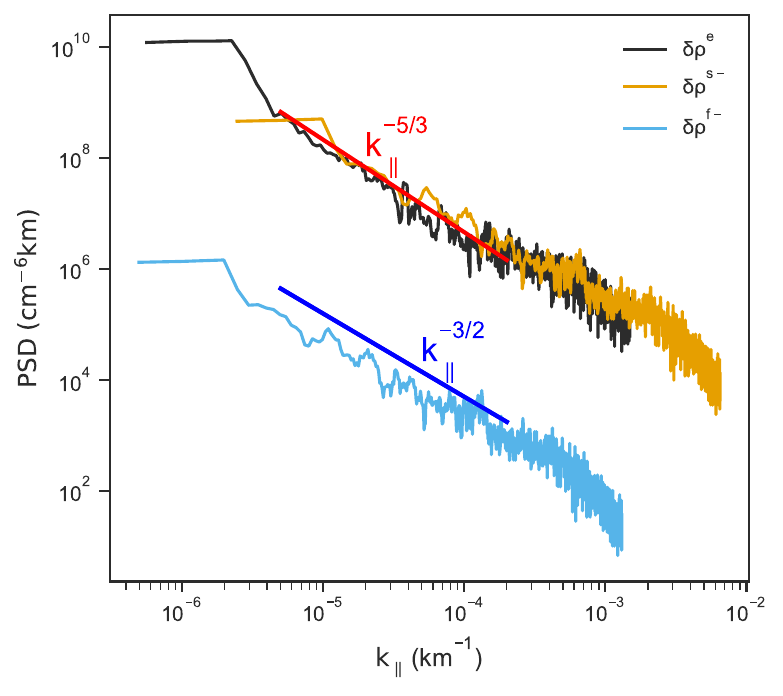}$$
	$$\includegraphics[width= 0.3\textwidth]{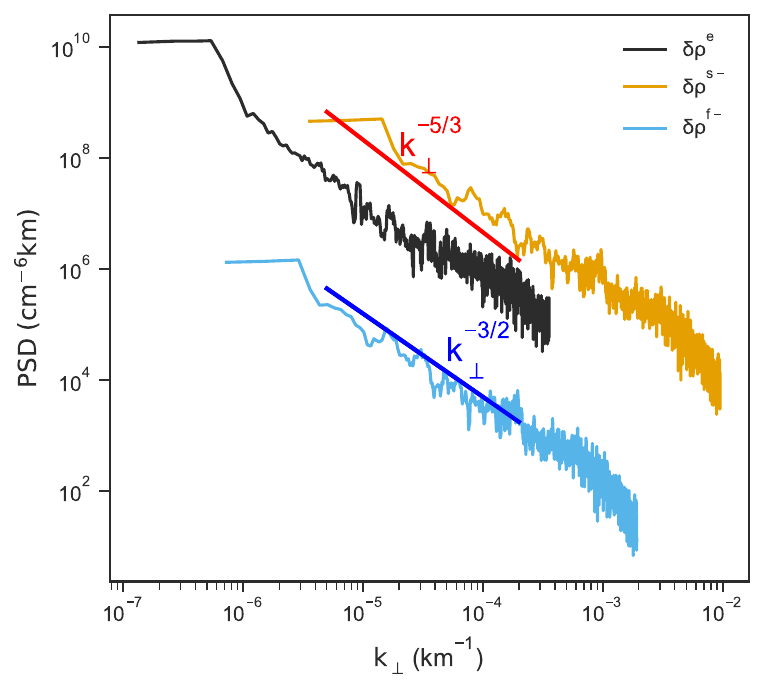}
	\includegraphics[width= 0.3\textwidth]{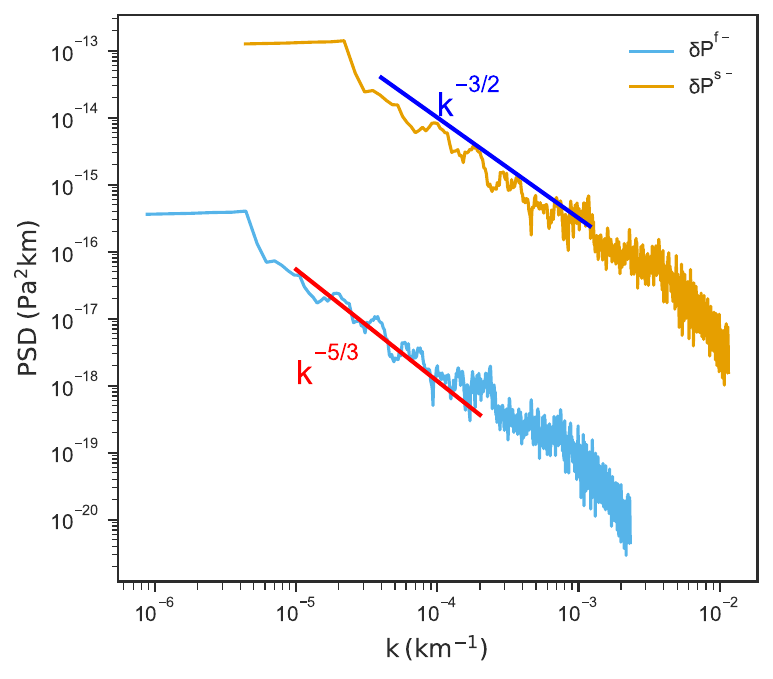}$$
	\caption{\small Power spectral density (PSDs) plots for the fluctuating density in frequency $f$ (top panels), wavenumber $k$ and parallel wavenumber $k_{\parallel}$ (middle panels), and perpendicular wavenumber $k_{\perp}$ (bottom left) and  pressure in wavenumber $k$ (bottom right). The top left panel shows the entropy (black curve) and forward slow (orange) and fast (blue) ms density PSDs, and the top right panel shows the entropy (black curve) and backward slow (orange) and fast (blue) ms density PSDs. The middle left and right panels show the entropy (black curve) and backward fast (blue) and backward slow (orange) ms density PSDs as a function of wavenumber $k$ and $k_{\parallel}$ respectively. The bottom left panel shows the corresponding PSDs as a function of $k_{\perp}$. The bottom right panel shows the fluctuating pressure PSD for the backward slow (orange) and fast (blue) ms modes. The red and blue lines show power laws with indices $-5/3$ and $-3/2$ respectively. } \label{fig:2}
\end{figure}

The three panels of Figure \ref{fig:1} show, from top to bottom, the time series of the normalized fluctuating density for the composite 10 30-minute subintervals associated with non-propagating i.e., advected entropy modes $\delta \rho^e/\rho_0$, forward and backward fast magnetosonic modes $\delta \rho^{f\pm}/\rho_0$, and forward and backward slow magnetosonic modes $\delta \rho^{s\pm}/\rho_0$. Figure 1 shows that the backward propagating fast ms mode contributes negligibly to the compressible density fluctuations in the sub-Alfv\'enic interval. The forward and backward slow ms modes are comparable although somewhat dominated by the backward mode. This is indeed borne out in Figure \ref{fig:2}, which plots the power spectral density (PSD) of the density fluctuations for the entropy and fast and slow forward and backward ms modes as a function of frequency $f$ (Hz) and wavenumber $k$ (km${}^{-1}$).\footnote{In the sub-Alfv\'enic interval of 2021 April 28, the magnetic field strength is approximately 315 nT, for which the proton cyclotron frequency is about 4.8 Hz. We need to use both the magnetic field data and the plasma data for the mode decomposition analysis. The resolution of the combined data set is 3.5 seconds, corresponding to a Nyquist frequency of 0.14 Hz, as shown in our spectra (see also \cite{Zank_etal_2022}). Owing to the limited resolution of the plasma data, the proton cyclotron frequency is therefore much higher than the frequency range we can consider. We consider MHD scales only, which have frequencies well below the proton cyclotron frequency ($\sim 4.8$ Hz).}
As noted above, the mode-decomposition, being linear, relates the observed frequency to the wavenumber through the appropriate dispersion relation \citep{Zank_etal_2023}, thereby avoiding the complications of Taylor's hypothesis in the sub-Alfv\'enic interval. For convenience, we list the transformations of frequency to wavenumber for the different wave modes at the end of the Appendix, listing the relevant equations from \cite{Zank_etal_2023}. 
Evidently, the entropy and the backward slow ms modes are the principle or dominant contributors to the density fluctuations in this particular parcel of sub-Alfv\'enic wind. The top two panels show that the frequency spectra for the entropic and magnetosonic modes are power laws in frequency. Power law curves for $f^{-5/3}$ and $f^{-3/2}$ in red and blue respectively are overplotted to guide the eye. The density variance associated with the entropy modes appears to be consistent with an $f^{-5/3}$ power law while the ms modes have a density variance that appears to be better described by a $f^{-3/2}$ curve. The middle left panel shows the density PSDs for the entropy and backward fast and backward slow ms modes in wavenumber, which closely resemble $k^{-5/3}$ (entropy and slow ms) and $k^{-3/2}$ power laws respectively. Of particular interest is the density PSD for the density fluctuations in light of Solar Orbiter Metis observations reported by \cite{Telloni_etal_2023a} describing the evolution of density fluctuations from 1.8 to 3 R${}_{\odot}$. \cite{Telloni_etal_2023a} find that the spectral exponent of the density PSD changes from $-2.32$ to $-1.64$ over this distance, regardless of whether the solar corona was observed in low- or high-density regions. The evolution of the density spectrum towards a Kolmogorov scaling was interpreted as the development of fully developed turbulence by about 3 R${}_{\odot}$. As discussed in \cite{Telloni_etal_2023a} and \cite{Zank_etal_2017a, Zank_etal_2020, Adhikari_etal_2023}, advected density fluctuations respond as a passive scalar to the velocity fluctuations, particularly the 2D incompressible component associated with Alfv\'en vortices, to form a Kolmogorov power law-like distribution in the density variance. 

The right middle and left bottom plots show the density variances of the same modes plotted as functions of parallel $k_{\parallel}$ and perpendicular $k_{\perp}$ wavenumber. Both the dominant entropy and backward slow ms density fluctuations follow $k_{\parallel}^{-5/3}$ and $k_{\perp}^{-5/3}$ power laws although both exhibit some flattening at higher $k$ values. Both panels indicate that the anisotropy of the dominant density fluctuations differs by mode, and this is discussed further in Section \ref{sec:4}.  

Since the ms pressure fluctuations $\delta p^{fs\pm}$ are typically proportional to the corresponding density fluctuations $\delta \rho^{fs\pm}$ \citep{Zank_Matthaeus_1992b,  zank_matthaeus_1993_incompress, Zank_etal_2017a, Zank_etal_2023} (although see \cite{Zank_etal_1990} for a more complicated equation of state), the time series are not plotted. However, in the bottom right panel of Figure \ref{fig:2}, we plot the wavenumber PSDs for the fluctuating ms pressures  $\delta p^{f-}$ and $\delta p^{s-}$, illustrating that the backward slow ms mode pressure spectrum resembles a $\sim k^{-3/2}$  and the backward fast mode pressure spectrum a $\sim k^{-5/3}$ power law. Curiously, the $-5/3$ density spectrum is mapped to a $-3/2$ pressure spectrum and vice versa. 

\begin{figure}
	$$\includegraphics[width= 0.5\textwidth]{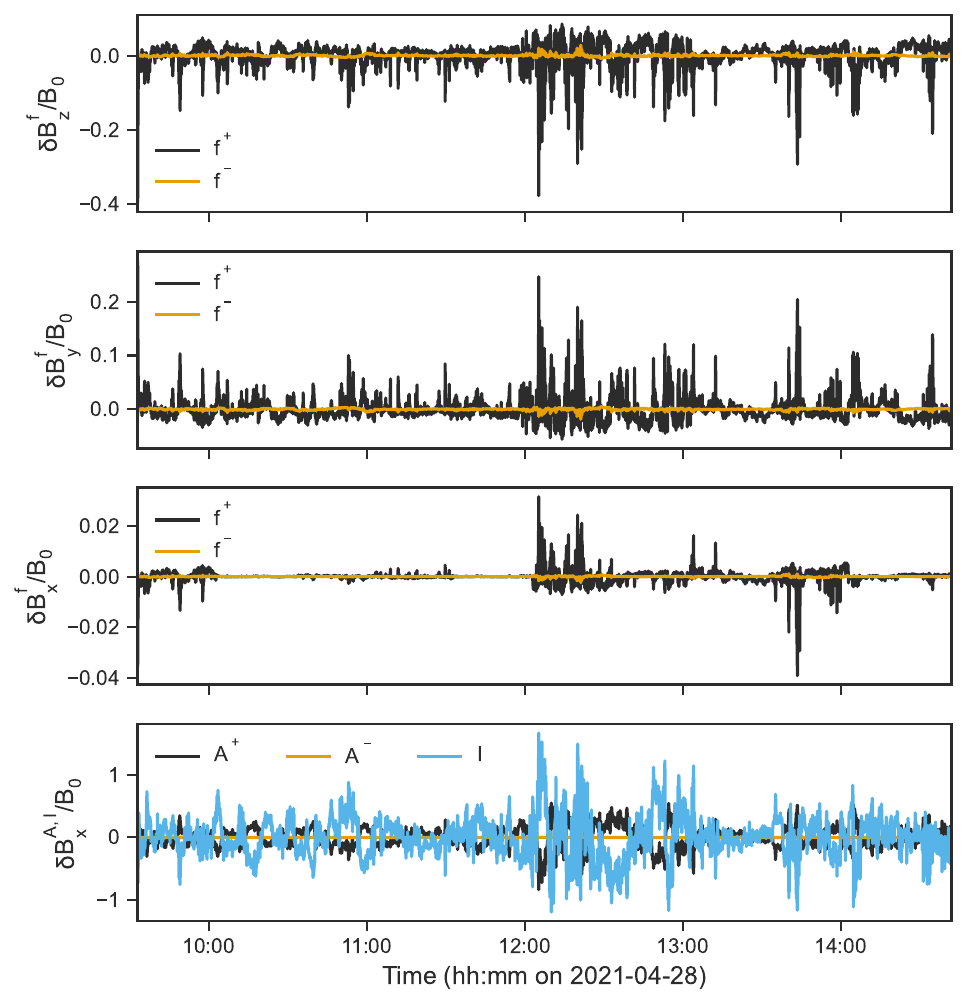}
	\includegraphics[width= 0.5\textwidth]{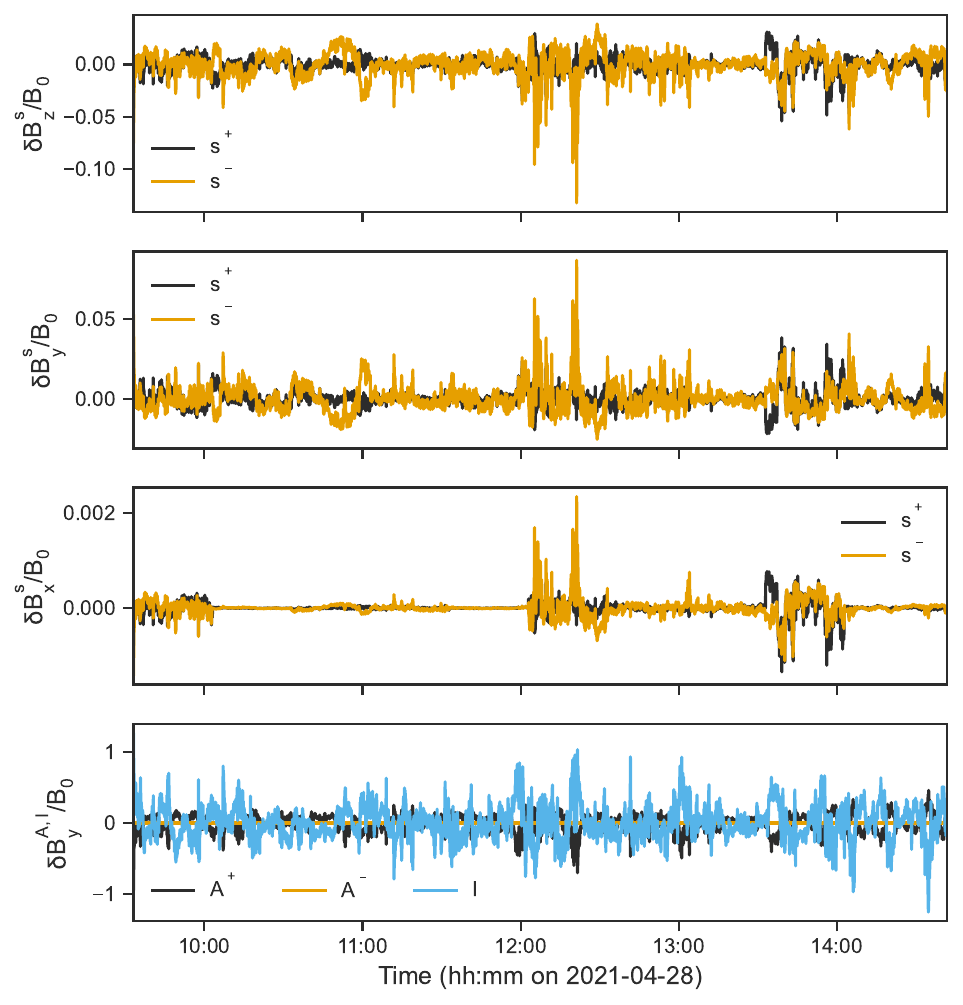}$$
	\caption{\small  Time series of the magnetic field fluctuations over the composite 10 30-minute subintervals of the 5-hour sub-Alfv\'enic interval showing normalized magnetic field fluctuations. {\bf Top left panel:} Parallel ($z$-component) magnetic field forward (black curve) and backward (orange curve) ($\pm$) fast magnetosonic (ms) modes $\delta  B_z^{f\pm}/B_0$. {\bf Middle left two panels:} forward and backward transverse magnetic field fluctuations $\delta B_{x,y}^{f\pm}$ for fast ms modes. {\bf Left bottom panel:} forward (black curve) and backward (orange curve) Alfv\'en ($\delta B_x^{A\pm}$) and advected magnetic island ($\delta B_x^i$, blue curve) modes. {\bf Top right panel:} Parallel ($z$-component) magnetic field forward (black curve) and backward (orange curve) ($\pm$) slow ms modes $\delta  B_z^{s\pm}/B_0$. {\bf Middle right two panels:} forward and backward transverse magnetic field fluctuations $\delta B_{x,y}^{s\pm}$ for slow ms modes. {\bf Left bottom panel:} forward (black curve) and backward (orange curve) Alfv\'en ($\delta B_y^{A\pm}$) and advected magnetic island (blue curve, $\delta B_y^i$) modes.  } 
	\label{fig:3}
\end{figure}

\begin{figure}
	$$\includegraphics[width= 0.3\textwidth]{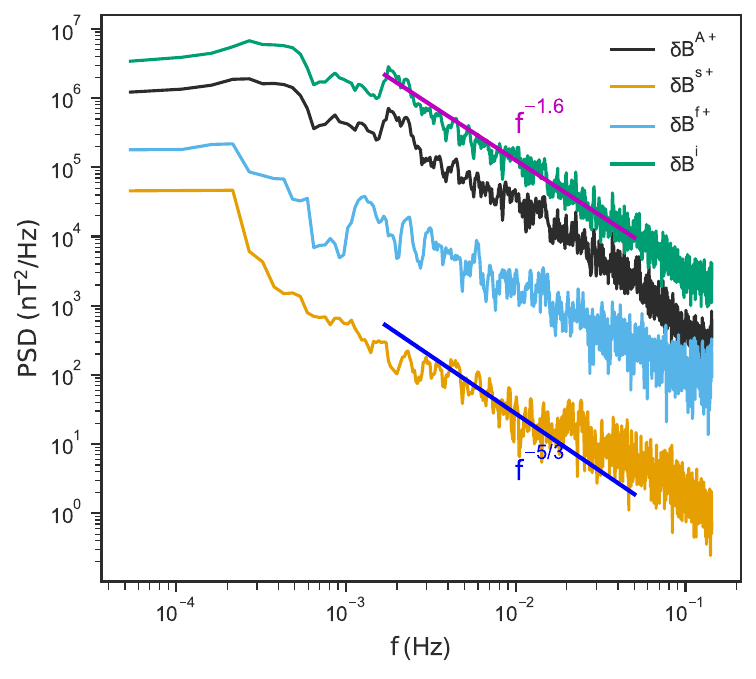}
	\includegraphics[width= 0.3\textwidth]{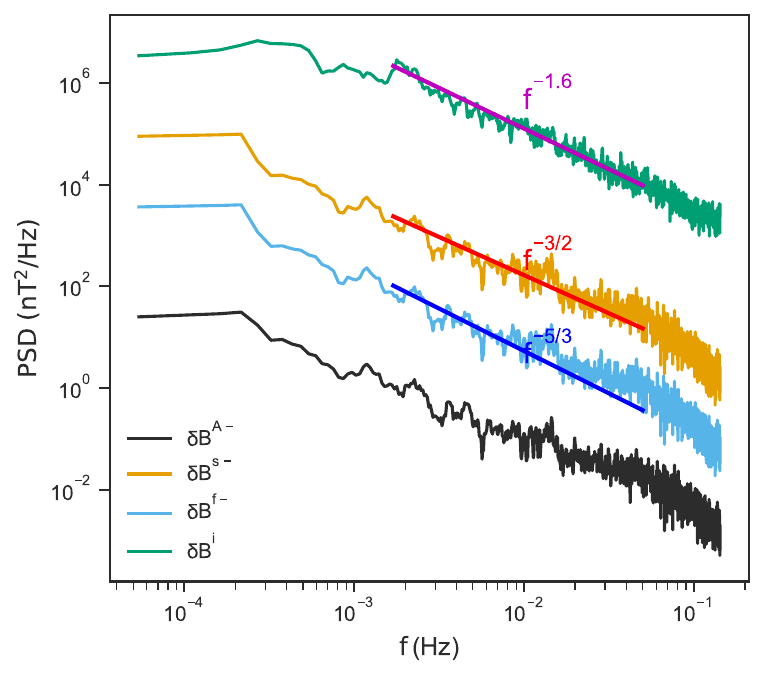}$$
	$$\includegraphics[width= 0.25\textwidth]{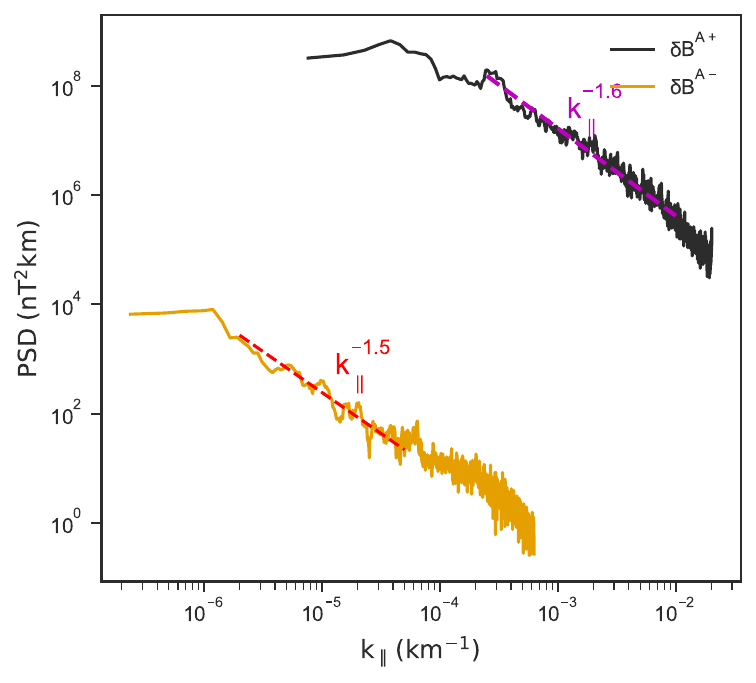}
	\includegraphics[width= 0.25\textwidth]{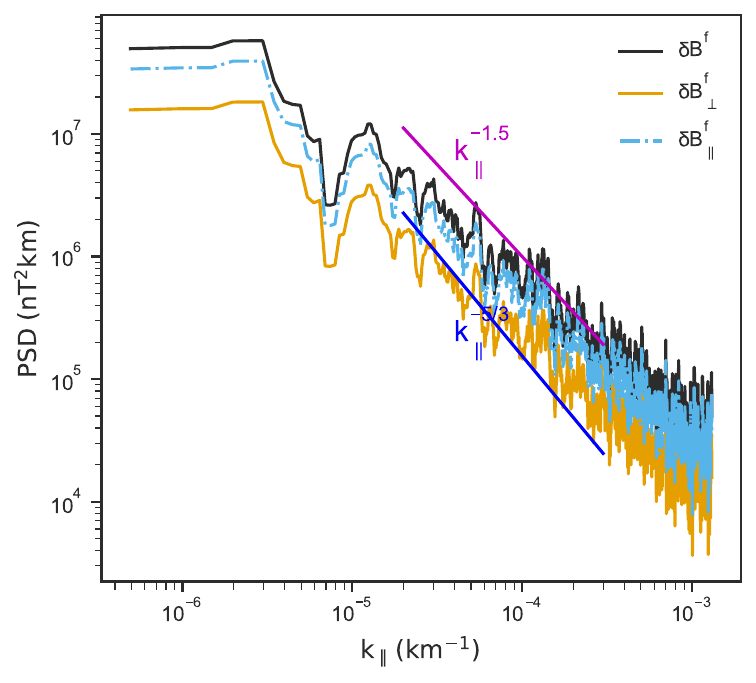}
	\includegraphics[width= 0.25\textwidth]{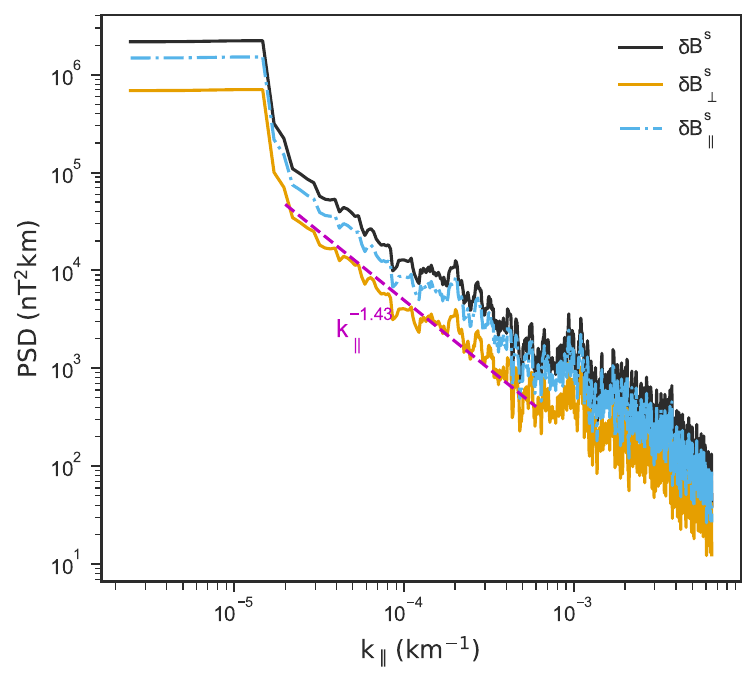}$$
	$$\includegraphics[width= 0.25\textwidth]{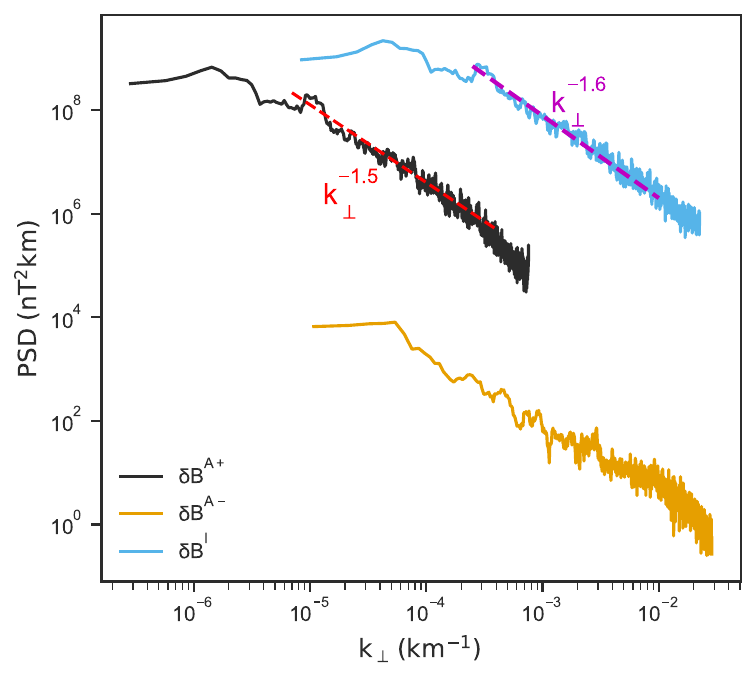}
	\includegraphics[width= 0.25\textwidth]{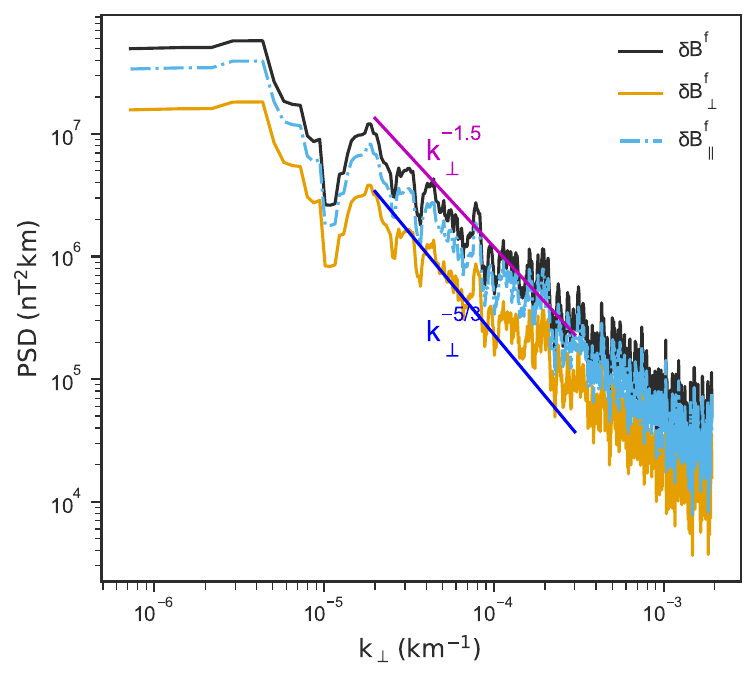}
	\includegraphics[width= 0.25\textwidth]{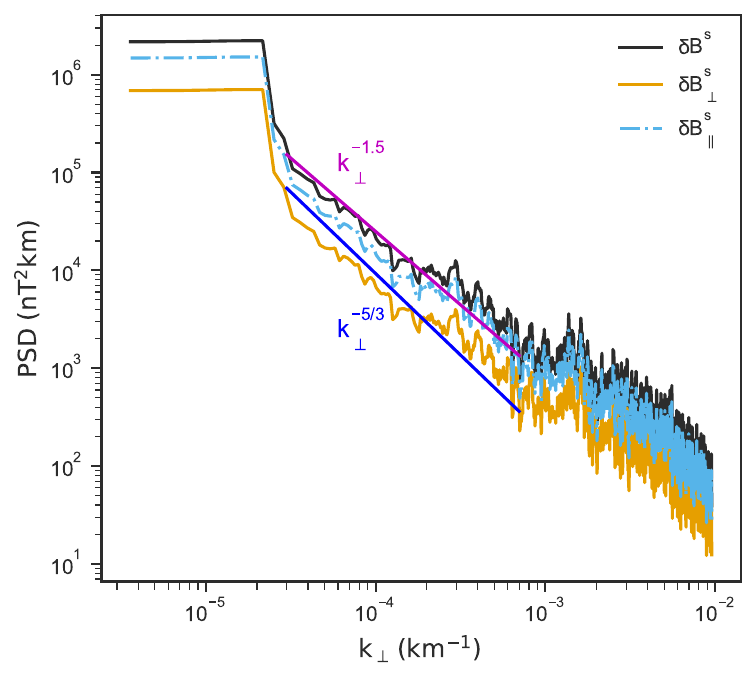}$$
	\caption{\small PSDs for the fluctuating magnetic field in frequency $f$ (top two panels), parallel wavenumber $k_{\parallel}$ (middle three panels), and perpendicular wavenumber $k_{\perp}$ (bottom three panels). The top left panel shows the magnetic island (green curve), the  forward Alfv\'en (black), forward fast (blue) and forward slow (orange) ms mode PSDs, and the top right panel shows the magnetic island (green curve), backward slow (orange) and fast (blue) ms and backward Alfv\'en mode PSDs. The middle panels show from left to right PSDs in $k_{\parallel}$ of the forward and backward Alfv\'en modes, the parallel $\delta B_{\parallel}^f$, transverse $\delta B_{\perp}^f$, and total $\delta B^f$ of the forward fast mode (the backward propagating fast mode is not shown - see Figure \ref{fig:3}), and the corresponding PSDs for the backward propagating slow ms mode. The bottom panels show from left to right PSDs in $k_{\perp}$ of the magnetic island and forward and backward Alfv\'en modes, the parallel $\delta B_{\parallel}^f$, transverse $\delta B_{\perp}^f$, and total $\delta B^f$ of the forward fast mode, and the corresponding quantities for the backward propagating slow ms mode.  } \label{fig:4}
\end{figure}

The fluctuating magnetic field data is illustrated in Figure \ref{fig:3} (the time series) and Figure \ref{fig:4} (spectral plots) for the components of the forward and backward fast ($\delta B_{x,y,z}^{f\pm}$) and slow ($\delta B_{x,y,z}^{s\pm}$) ms modes, the magnetic island ($\delta B_{x,y}^i$) modes, and the forward and backward Alfv\'en ($B_{x,y}^{A\pm}$) waves. The magnetic island and Alfv\'enic modes possess of course only incompressible transverse magnetic fluctuations while the ms modes possess both transverse and longitudinal components. Figure \ref{fig:3} illustrates again that the backward fast ms mode contributes negligibly to the compressible fluctuations, which are comprised primarily of the forward and backward slow and the forward fast magnetosonic modes. The incompressible transverse fluctuations are the dominant contribution to the magnetic field energy density. Although apparent in Figure \ref{fig:3}, this is particularly well illustrated in the PSD plots of Figure \ref{fig:4}. The top panels show PSDs as a function of $f$ for all magnetic modes, the middle panels as a function of the parallel wavenumber $k_{\parallel}$ (where ${\bf k} = ({\bf k}_{\perp}, k_{\parallel})$) for all but the magnetic island modes, and the bottom panels as a function of $k_{\perp} = |{\bf k}_{\perp}|$ for all magnetized modes. To guide the eye, we plot red and blue lines corresponding to power laws with various indices ($-1.6$, $-3/2$ and $-5/3$). 

The magnetic island mode is evidently the dominant population (Figure \ref{fig:4}, bottom left), followed by the forward Alfv\'en modes. From the top two panels, the magnetic island frequency spectrum is consistent with a power law of the form $f^{-1.6}$, which appears to be quite similar to the backward Alfv\'en modes, whereas the forward Alfv\'en modes appear to have a marginally flatter spectrum, perhaps more consistent with a $f^{-3/2}$ spectrum in frequency. 
The ms modes appear to have spectra slightly flatter than $f^{-3/2}$. As shown in the left middle and bottom plots of Figure \ref{fig:4}, the forward and backward Alfv\'en modes can be described by power laws in the parallel and perpendicular wavenumbers as $k_{\parallel}^{-1.6}$, $k_{\perp}^{-1.5}$ (forward) and $k_{\parallel}^{-1.5}$, $k_{\perp}^{-1.5}$ (backward) respectively. The 
magnetic island spectrum has a power law $\sim k_{\perp}^{-1.6}$ (bottom left panel), which, since it is an advected mode, is consistent with the corresponding frequency PSD. The middle and rightmost panels of the middle and bottom rows of Figure \ref{fig:4} plot the magnetic PSDs  of the forward fast and backward slow ms modes as functions of $k_{\parallel}$ and $k_{\perp}$ respectively (the two dominant compressible modes). 
The fast and slow ms magnetic spectra have a modestly more significant longitudinal component than transverse component and the spectral amplitude per logarithmic wavenumber ($\delta B_{\parallel}^2 {\rm d}k_{\parallel, \perp}$) of the fast ms mode is at least an order of magnitude larger than that of the slow ms mode. This is due to the fast magnetosonic mode being primarily magnetic and the slow ms mode being essentially a sound wave in a low beta plasma. 
The spectra of the longitudinal fluctuations are flatter than $k^{-3/2}$ (fits suggest a range of power law indices from $-1.41$ to $-1.45$ in $k_{\parallel}$). The transverse fluctuations of the fast ms modes appear to have a spectral index between $-3/2$ and $-5/3$ whereas those of the slow ms mode have a very hard spectrum. The fast and slow ms spectra in $k_{\perp}$, at least for the longitudinal component, has a power law index of $\sim -3/2$, whereas the transverse fluctuations may possess a slightly steeper spectrum, possible $\sim k_{\perp}^{-5/3}$. 
The amplitudes of the compressible ms magnetic PSD plots are significantly lower than those of the incompressible magnetic PSD plots associated with both the magnetic islands and the forward Alfv\'enic modes. Specifically, the amplitude of $\delta B^2 dk_{\perp,\parallel}$ is at least an 
order of magnitude greater for the magnetic island and forward Alfv\'en PSDs than the dominant forward fast ms mode PSD. The incompressible magnetic field modes are therefore the dominant component compared to the compressible component \citep[see also][]{Zhao_etal_2023}, indicating that the 5-hour plasma parcel is essentially incompressible. 

\begin{figure}
	$$\includegraphics[width= 0.5\textwidth]{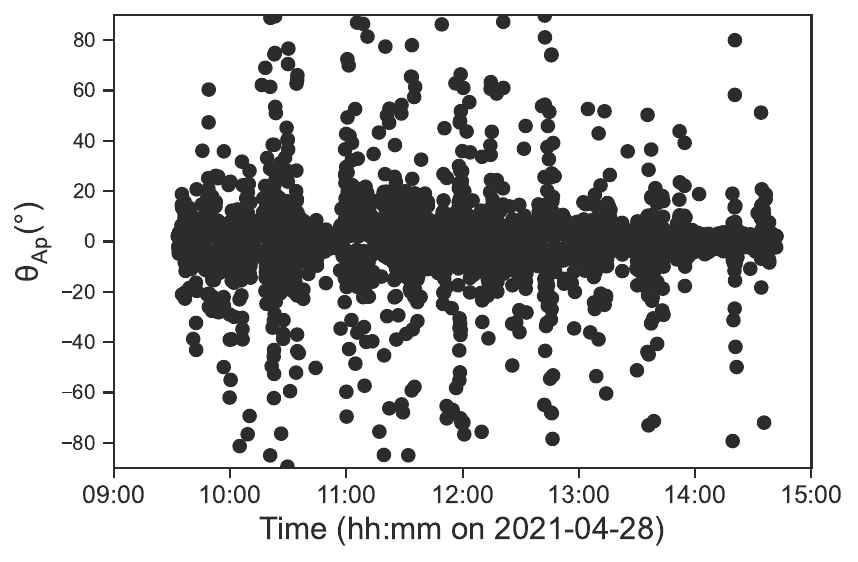}
	\includegraphics[width= 0.5\textwidth]{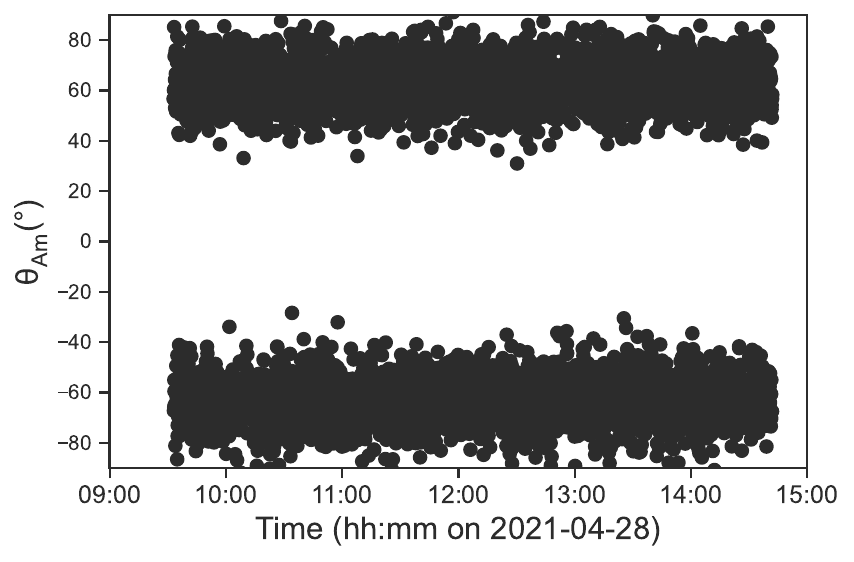}$$
	\caption{\small {\bf Left:} Plot of the phase angle $\theta^{A+}$ for each of the forward Alfv\'en modes over the full 5-hour interval based on the set of 10 30-minute subintervals. {\bf Right:} Plot of the phase angle $\theta^{A-}$ for each of the backward Alfv\'en modes over the 5-hour interval. } \label{fig:4a}
\end{figure}

Unlike the anisotropy results of \cite{Bandyopadhyay_McComas_2021, Zhao_etal_2022}, who found  that the 2D:slab ratio of magnetic energies is $\sim 0.43$ between 27.95 -- 64.5 R${}_{\odot}$, the mode-decomposition analysis indicates that in this sub-Alfv\'enic interval, 2D magnetic islands are  the dominant component of the transverse (and total) magnetic field fluctuations rather than slab-like Alfv\'enic fluctuations. The energy density of the magnetic island component $\langle \delta {B^i}^2 \rangle$ in the 5-hour interval is found by integrating under the spectrum, being $\langle \delta {B^i}^2 \rangle = 1\times 10^6$ nT${}^2$. By contrast, the energy density of the Alfv\'en component $\langle \delta {B^A}^2 \rangle$ in the 5-hour interval is $\langle \delta {B^A}^2 \rangle = 2.3 \times 10^5$ nT${}^2$, which yields a variance anisotropy of $\langle \delta {B^i}^2 \rangle / \langle \delta {B^A}^2 \rangle = 4.1$. This is the same as found by \cite{Bieber_etal_1996JGR}, i.e., a ratio of 4:1, for the supersonic solar wind at 1 au. Obviously, variance anisotropy found by \cite{Bieber_etal_1996JGR} was derived on the basis of a quite different set of methods and assumptions, most notably that there was no distinction between transverse component contributions from compressible and incompressible modes. The result presented here suggests that indeed the plasma at 1 au measured in the \cite{Bieber_etal_1996JGR} analysis was essentially incompressible. 
As a nominal estimate of the energy density in 2D versus slab fluctuations, this is in agreement with the ratio predicted of the variance anisotropy for NI MHD in the $\beta_p \ll 1$ or $O(1)$ regimes \citep{Zank_etal_2020}.  

Besides magnetic islands, highly oblique Alfv\'enic fluctuations are essentially quasi-2D with $k_{\parallel} \ll k_{\perp}$ and are effectively non-propagating (see the discussion in the Appendix of \cite{Zank_etal_2017a}). Such highly oblique  Alfv\'enic modes or Alfv\'en vortices form a component of the leading order $\beta_p \ll 1$ or $O(1)$ nearly incompressible MHD description \citep{Zank_etal_2017a} and introduce  quasi-2D velocity fluctuations. Since we can calculate the phases of the Alfv\'en fluctuations via the mode-decomposition, we plot in Figure \ref{fig:4a} the values of $\theta^{A\pm}$. The left plot shows $\theta^{A+}$, i.e., the obliquity of the forward propagating Alfv\'en modes over the full 5-hour interval, illustrating that the waves are propagating essentially along the background magnetic field (i.e., based on the 30-minute subinterval backgrounds into which the full sub-Alfv\'enic flow is decomposed), and are clustered roughly in the interval $-25^{\circ} \leq \theta^{A+} \leq 25^{\circ}$, with some highly oblique Alfv\'en modes or Alfv\'en vortices. By contrast, the minority backward propagating Alfv\'en modes, as illustrated in the right panel of Figure \ref{fig:4a}, exhibit a pronounced bimodal distribution in $\theta^{A-}$ such that $40^{\circ} \leq \theta^{A-} \leq 90^{\circ}$ or $-90^{\circ} \leq \theta^{A-} \leq -40^{\circ}$. On taking $ \theta^{A-} = 65^{\circ}$ as the median value, one obtains   $k_{\perp}^{A-} /k_{\parallel}^{A-} = 2.15$. Consequently, the backward propagating Alfv\'en waves are highly oblique in this sub-Alfv\'enic flow with those fluctuations possessing $|\theta^{A-}| \geq 65^{\circ}$ being weakly propagating or advected Alfv\'en vortices. 

The slab component is composed almost exclusively of uni-directionally propagating forward Alfv\'en waves. Such unidirectionally propagating Alfv\'en waves should not therefore exhibit a power law spectrum since nonlinear interactions require counter-propagating Alfv\'en waves \citep{Shebalin_etal_1983}. The formation of a power law for slab turbulence with a normalized cross helicity $|\sigma_c| \sim 1$ has been addressed in detail by \cite{Zank_etal_2020} who argued that ``scattering'' and passive advection by the dominant 2D component (perhaps better described as a generalization of sweeping, as discussed in \cite{Zhao_etal_2023}) would yield a power law spectrum. See also \cite{Alberti_etal_2022} for a related discussion and somewhat different interpretation.

\begin{figure}[!t]
	$$\includegraphics[width= 0.5\textwidth]{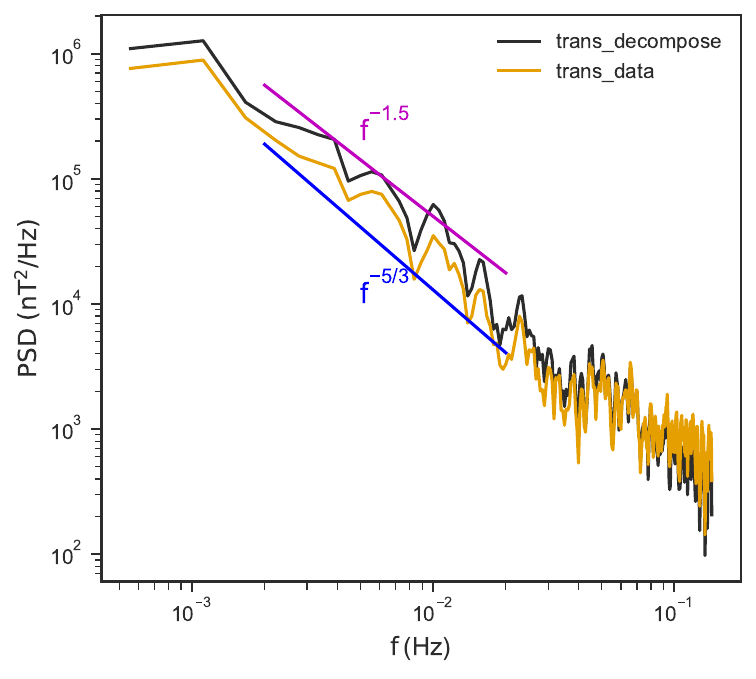}
	\includegraphics[width= 0.5\textwidth]{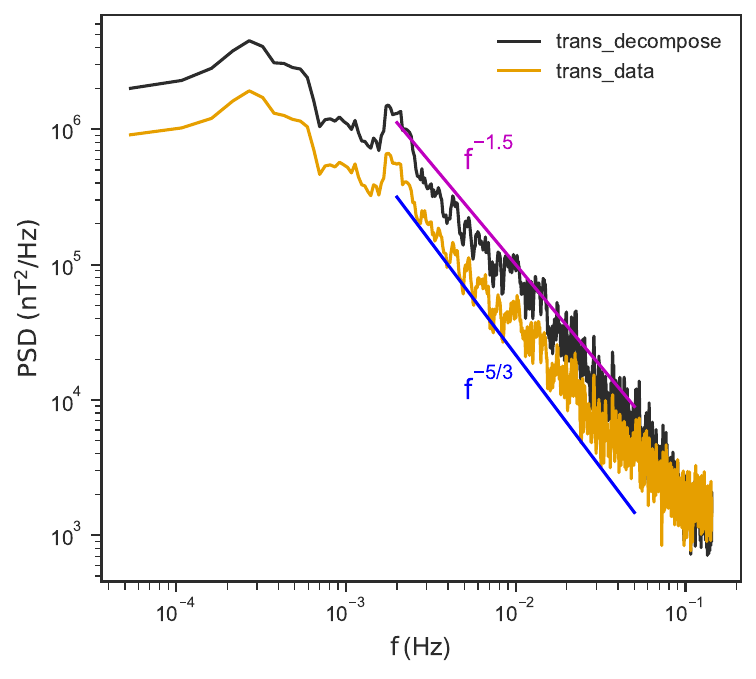}$$
	\caption{\small  {\bf Left:}  The orange curve shows the Fourier transform-derived PSD for the transverse magnetic field fluctuations for a 30-minute sub-interval from 10:33 to 11:03 UT on 2021-04-28.  A reconstructed spectrum for the transverse fluctuations derived from the mode-decomposition of this 30-minute subinterval is shown by the black curve, i.e., the summation of the separate  magnetic island, Alfv\'enic, and fast and slow ms transverse magnetic field contributions. {\bf Right:} The orange curve shows the Fourier transform-derived PSD for the transverse magnetic field fluctuations for the original 5-hour data interval. A reconstructed spectrum for the transverse fluctuations derived from the mode-decomposition of the 10 30-minute subintervals is shown by the black curve. } \label{fig:5}
\end{figure}

A measure of the accuracy of the linearized decomposition can be gleaned from a comparison of the frequency spectrum of the transverse magnetic fluctuations derived from a standard spectral analysis of the original data with a spectrum derived from the summation of the mode-decomposed transverse magnetic fluctuations i.e., the transverse magnetic fluctuations contributed by the magnetic island, Alfv\'en, and fast and slow ms modes \citep{Zank_etal_2023}. 
Illustrated in the left panel of Figure \ref{fig:5} is a comparison of the Fourier transform-derived frequency (orange line) and the mode-decomposition-constructed PSD (black line) for transverse magnetic fluctuations for a typical 30-minute subinterval. Evidently, the full nonlinear PSD and the mode-decomposition reconstructed PSD  follow each other very closely for the 30-minute intervals. However, an interesting question arises when combining the spectral results from the 10 30-minute subintervals into a single spectrum for the full 5 hour interval of interest, as we have done in presenting the density and the magnetic field fluctuations spectra in Figures \ref{fig:2} and \ref{fig:4}. In combining the spectral data from each of the subintervals, we have assumed implicitly that each of the subintervals provides an independent statistical realization of an (almost) identical system because the mean state of each subinterval does not differ significantly from any other. The recombination then corresponds to an ensemble average, in this case constructed from the 10 30-minute subintervals. The reassembled transverse magnetic field fluctuation frequency PSD using the spectral data from all 10 30-minute subintervals (black curve) is compared to the Fourier-transformed spectrum derived from the full 5-hour interval (orange curve) in Figure \ref{fig:5}, right panel. 
By restricting our attention to the frequency $f$, we avoid the complications of converting to wavenumber space. Relatively modest differences in spectral amplitude are present in the ``ensemble averaged'' mode-decomposed spectrum and the high-frequency part of the mode-decomposed spectrum steepens rather than flattens as in the Fourier frequency spectrum.  Nonetheless, the basic features of the two spectra are very similar. The general conclusion is that the linear mode-decomposition captures the basic spectral characteristics of the fully nonlinear individual 30-minute intervals very well, and since the background states of each of the subintervals are similar, the ``ensemble averaged'' PSD for the full 5-hour interval also captures the properly nonlinear spectrum rather well. As with the prior spectral plots, we superimpose two power law curves, $f^{-3/2}$ and $f^{-5/3}$ to guide the eye. Finally, we repeat that 
mode-decomposition does not represent a model for the behavior of fluctuations in the sub-Alfv\'enic flow but is simply a snapshot of the fundamental MHD modes during a time short enough that nonlinearity is unimportant. 

\begin{figure}
	$$\includegraphics[width= 0.5\textwidth]{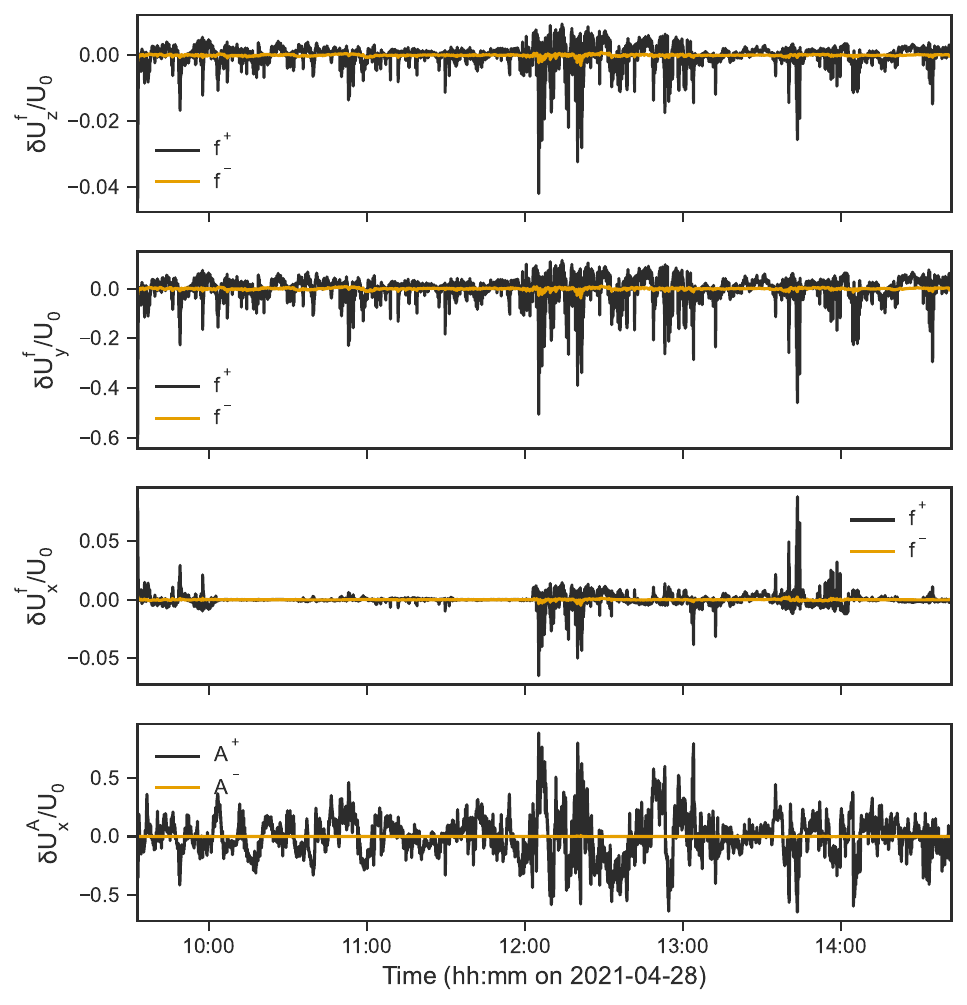}
	\includegraphics[width= 0.5\textwidth]{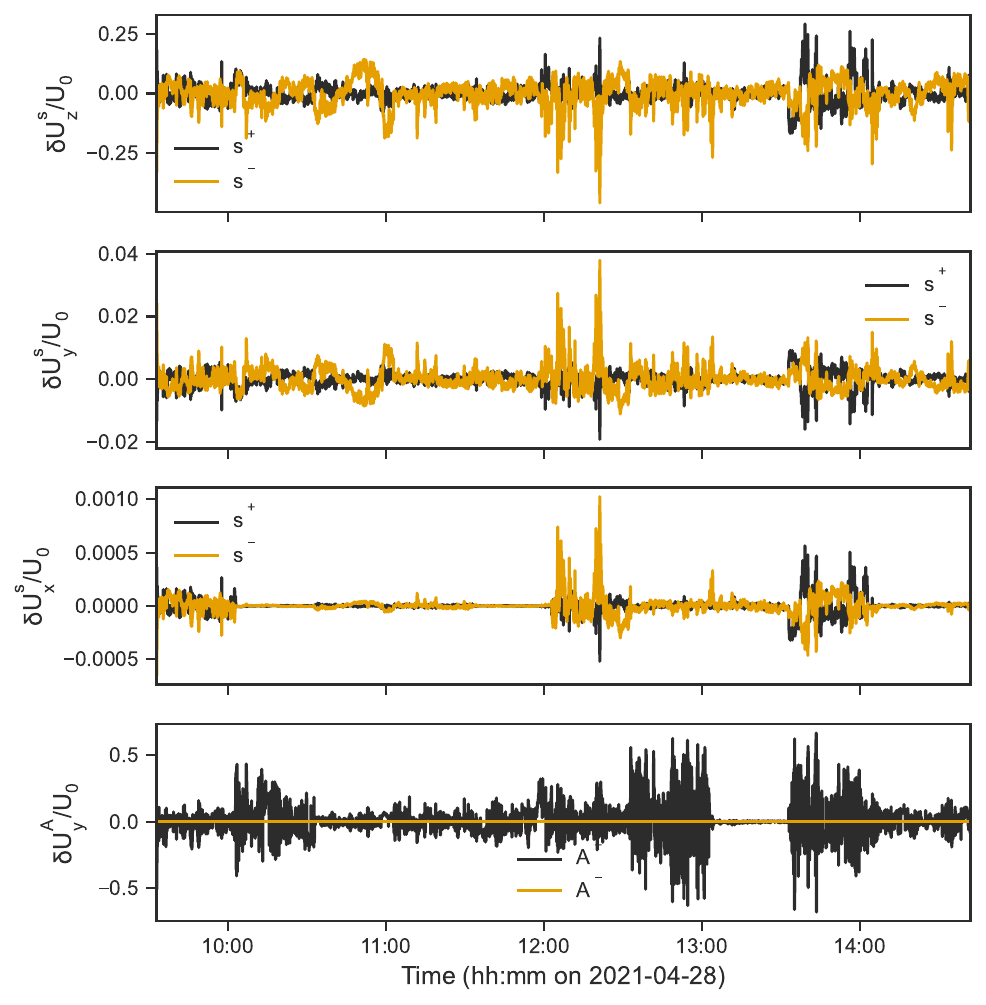}$$
	\caption{\small  The same format as Figure \ref{fig:3} for the time series of the normalized velocity field fluctuations. {\bf Top left panel:} Parallel ($z$-) velocity component for forward (black curve) and backward (orange curve) ($\pm$) fast ms modes $\delta  u_z^{f\pm}/U_0$. {\bf Middle left two panels:} forward and backward transverse velocity fluctuations $\delta u_{x,y}^{f\pm}$ for fast ms  modes. {\bf Left bottom panel:} forward (black curve) and backward (orange curve) Alfv\'en ($\delta u_x^{A\pm}$) modes. {\bf Top right panel:} Parallel ($z$-) velocity component forward (black curve) and backward (orange curve) ($\pm$) slow ms modes $\delta  u_z^{s\pm}/U_0$. {\bf Middle right two panels:} forward and backward transverse velocity fluctuations $\delta u_{x,y}^{s\pm}$ for slow ms  modes. {\bf Left bottom panel:} forward (black curve) and backward (orange curve) Alfv\'en ($\delta u_y^{A\pm}$) modes.  } \label{fig:6}
\end{figure}

\begin{figure}
	$$\includegraphics[width=4.5cm, height=3.3cm]{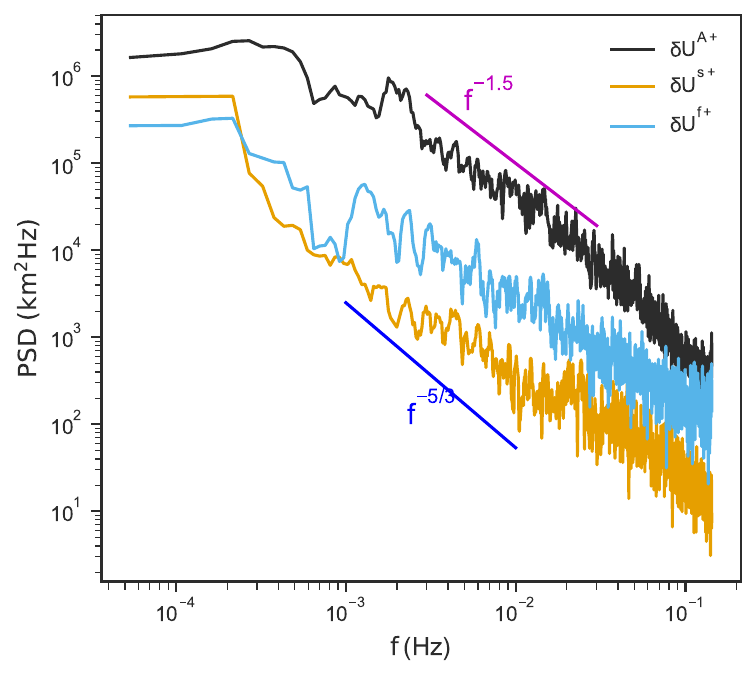}
	\includegraphics[width=4.5cm, height=3.3cm]{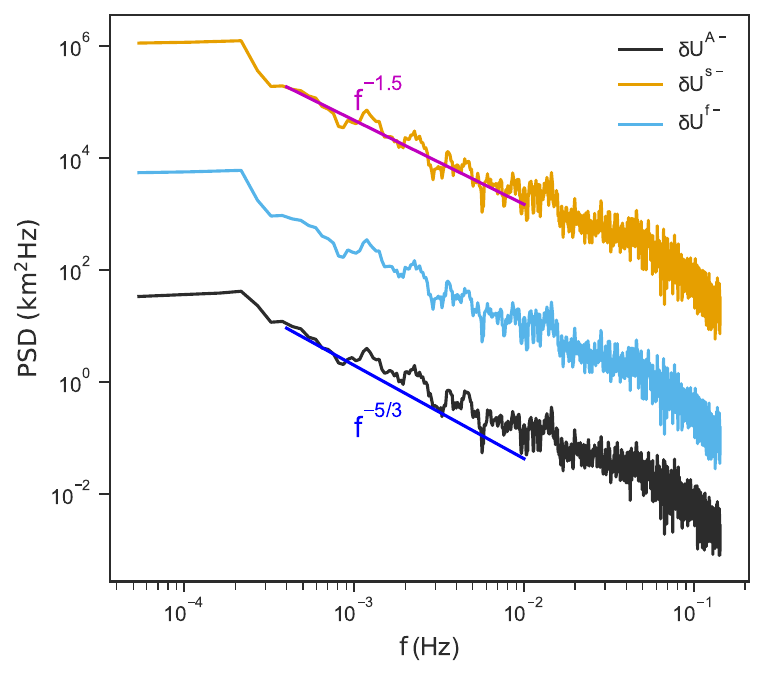} 
	\includegraphics[width=4.5cm, height=3.3cm]{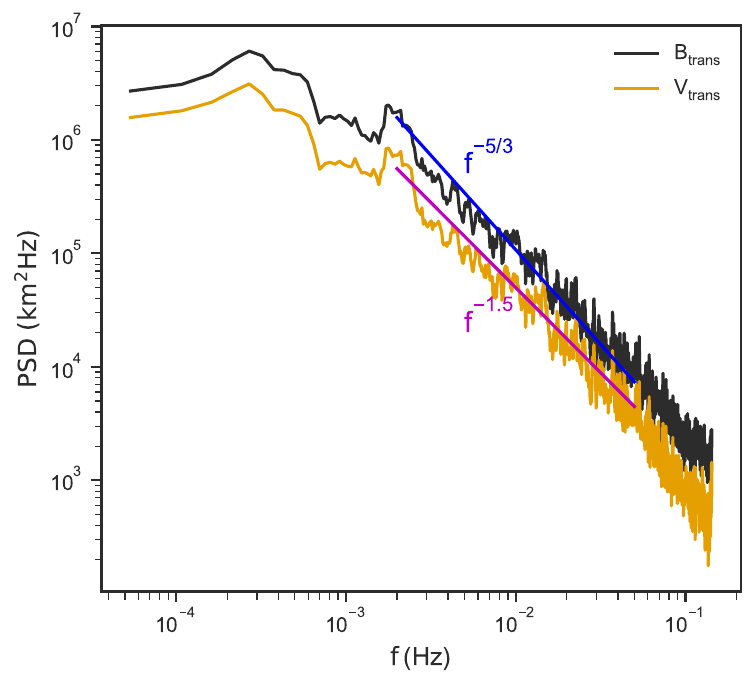}$$
	$$\includegraphics[width=4.5cm, height=3.3cm]{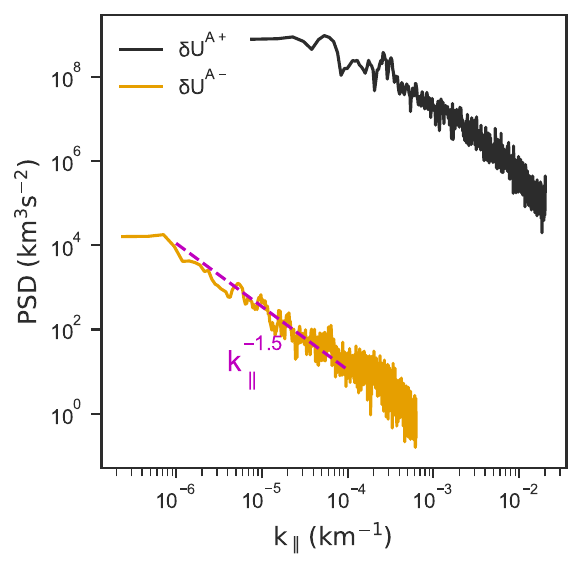}
	\includegraphics[width=4.5cm, height=3.3cm]{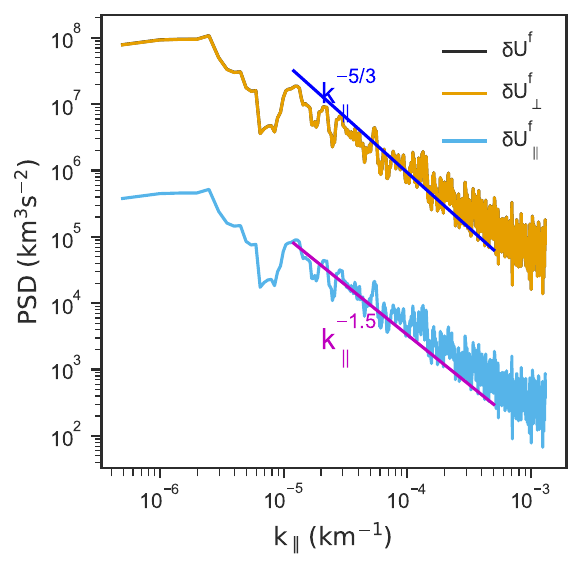}
	\includegraphics[width=4.5cm, height=3.3cm]{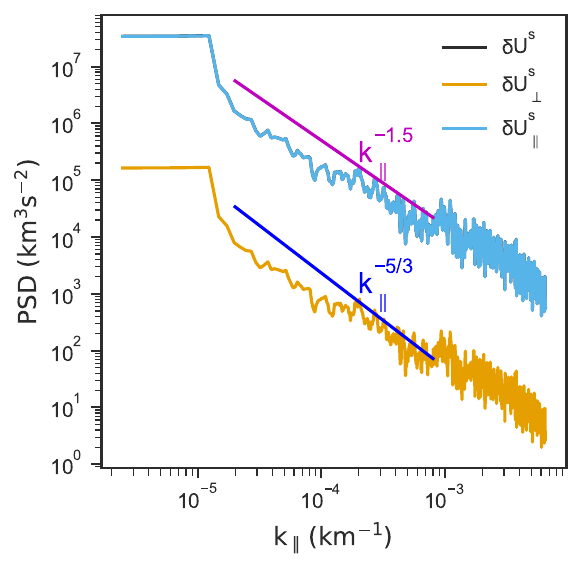}$$
	$$\includegraphics[width=4.5cm, height=3.3cm]{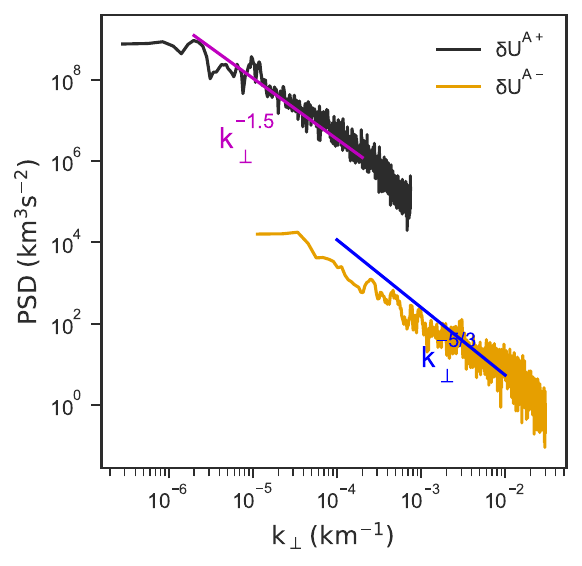}
	\includegraphics[width=4.5cm, height=3.3cm]{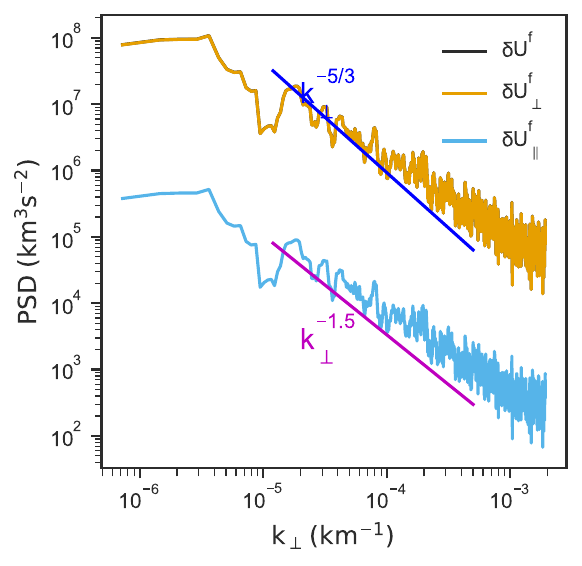}
	\includegraphics[width=4.5cm, height=3.3cm]{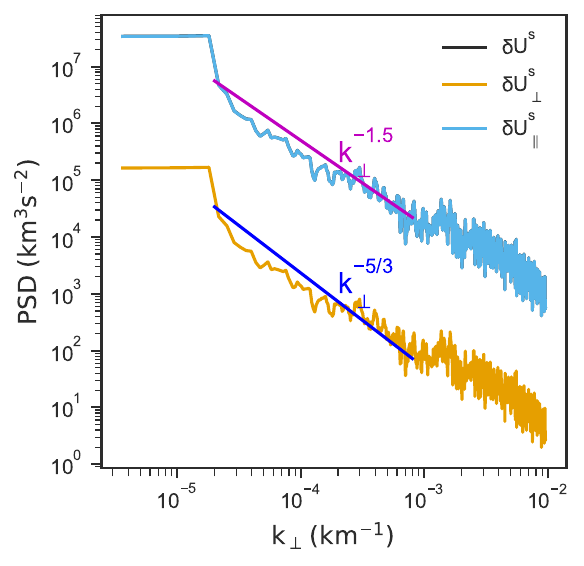}$$
	\caption{\small  PSDs for the fluctuating velocity in frequency $f$ (top three panels), parallel wavenumber $k_{\parallel}$ (middle three panels), and perpendicular wavenumber $k_{\perp}$ (bottom three panels). The top left panel shows the  forward Alfv\'en (black), forward fast (blue) and forward slow (orange) ms mode PSDs,  the top middle panel shows the  backward slow (orange) and fast (blue) ms and backward Alfv\'en mode PSDs. The top right panel shows a plot of the fluctuating transverse magnetic field PSD in Alfv\'en units (black curve) reconstructed from the transverse magnetic fluctuations obtained from the separate  magnetic island, Alfv\'enic, and fast and slow ms transverse magnetic field contributions. The orange curve shows a corresponding kinetic energy PSD reconstructed from the transverse velocity fluctuations obtained from the mode-decomposition for the separate Alfv\'enic, fast and slow ms transverse magnetic field contributions. The red and blue  lines corresponds to $f^{-3/2}$, $k^{-3/2}$ and $f^{-5/3}$, $k^{-5/3}$ power laws, respectively. 
		The middle panels show from left to right PSDs in $k_{\parallel}$ of the forward and backward Alfv\'en modes, the parallel $\delta u_{\parallel}^f$, transverse $\delta u_{\perp}^f$, and total $\delta u^f$ of the forward fast mode, and the corresponding PSDs for the backward propagating slow ms mode. The bottom panels show from left to right PSDs in $k_{\perp}$ of the forward and backward Alfv\'en modes, the parallel $\delta u_{\parallel}^f$, transverse $\delta u_{\perp}^f$, and total $\delta u^f$ of the forward fast mode , and the corresponding PSDs for the backward propagating slow ms mode.  } \label{fig:7}
\end{figure}

The decomposition of the velocity fluctuations is illustrated in Figure \ref{fig:6}. These  comprise the parallel $\delta u_z^{fs\pm}$ and transverse $\delta u_{x,y}^{fs\pm}$ velocity fluctuations of the forward ($+$) and backward ($-$) fast ($f$) and slow ($s$) ms modes, and the transverse $\delta u_{x,y}^{A\pm}$ velocity fluctuations of the forward and backward Alfv\'enic modes. The fast ms modes are  primarily forward propagating and, unlike the corresponding magnetic field fluctuations, are dominated by the transverse velocity component. By contrast, the parallel fluctuating component dominates the velocity fluctuations of the backward and forward slow ms mode. The dominance of the fluctuating transverse velocity component for the fast ms mode versus the dominance of the fluctuating parallel component is evident in the wavenumber spectra illustrated in Figure \ref{fig:7}. Both sets of ms spectra exhibit power laws in $k_{\parallel}$ and $k_{\perp}$ with very flat spectra,  both having a power law exponent of $\sim -1.4$.  In this, they differ from the incompressible transverse velocity fluctuations associated with the Alfv\'en modes. As with the magnetic field fluctuations, Figures \ref{fig:6} and \ref{fig:7} show that the forward Alfv\'en modes is the dominant Alfv\'en mode, i.e., the Alfv\'enic component is essentially uni-directional and thus highly anisotropic with a slab cross helicity $|\sigma_c^A| \sim 1$. The contribution to the kinetic energy spectra in both $k_{\parallel}$ and $k_{\perp}$ by the compressible transverse velocity fluctuations of the forward fast  ms modes is comparable to the incompressible transverse velocity fluctuations from the Alfv\'en modes. Secondly, the ms and Alfv\'enic spectral slopes differ somewhat, as can be seen from the  $ k_{\parallel, \perp}^{-1.5}$ and $k_{\parallel, \perp}^{-5/3}$ curves superimposed over the various spectra. If nothing else, this result suggests that, unlike the fluctuating magnetic field, the transverse velocity fluctuations may be comprised equally of incompressible and compressible fluctuations, both of which exhibit slightly different spectral forms in wavenumber. Consequently, this may result in a transverse velocity variance or kinetic energy wavenumber spectrum that looks rather different from the fluctuating transverse magnetic field variance wavenumber spectrum that is dominated by incompressible transverse modes. Despite these very evident differences in the wavenumber spectra, the plots in the top right panel of Figure \ref{fig:7} showing (black curve) the frequency PSD of the transverse fluctuating magnetic field expressed in Alfv\'en units and the kinetic energy PSD for transverse velocity fluctuations frequency PSD (orange curve) differ only in amplitude, being a factor of about two different, but are otherwise essentially the same with the same power law index ($\sim -1.5$). The magnetic and kinetic energy PSDs were
constructed from the transverse magnetic and  velocity components extracted from the mode decomposition of the magnetic island modes (magnetic PSD only), Alfv\'en, and fast and slow ms modes. 

\begin{figure}
	$$\includegraphics[width= 0.5\textwidth]{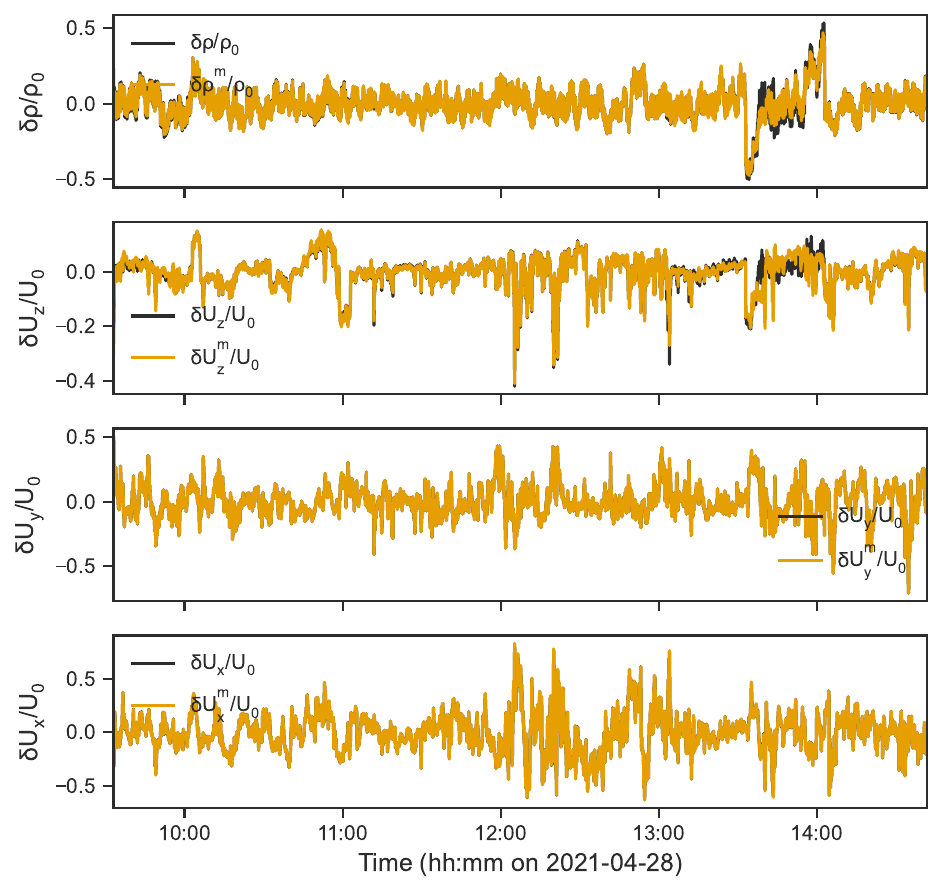}
	\includegraphics[width= 0.5\textwidth]{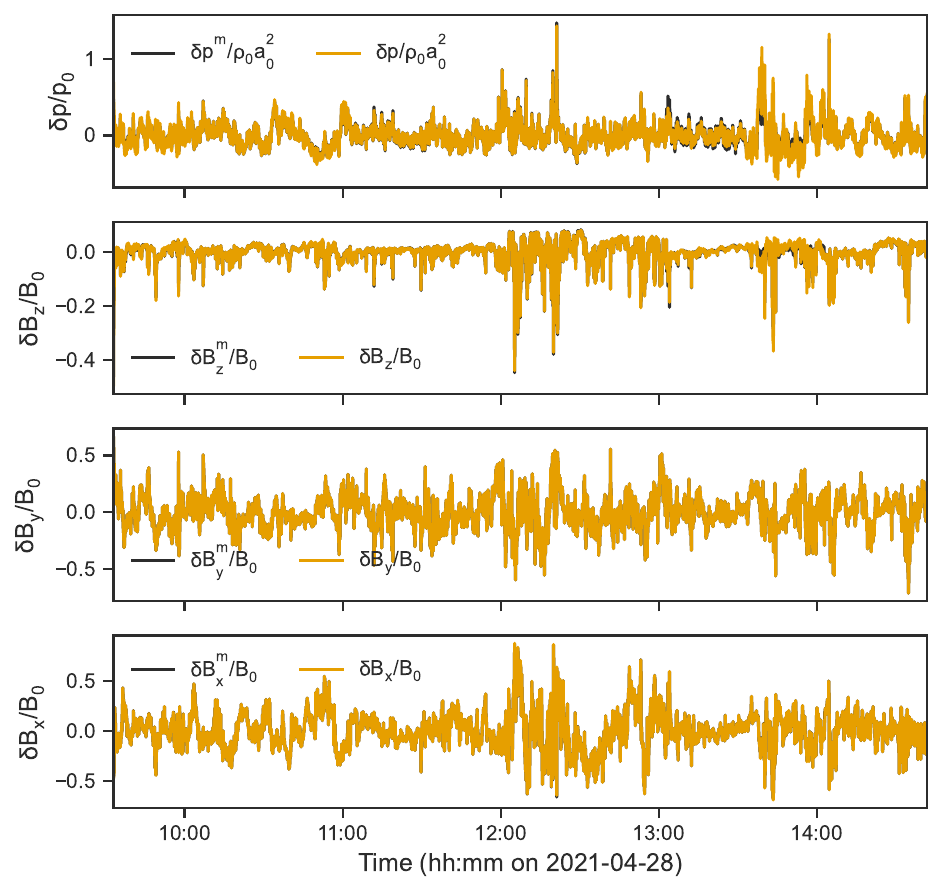}$$
	\caption{\small  The normalized measured data for the full 5-hour sub-Alfv\'enic interval, plotted in orange, overplotted  with the reconstructed values derived from the ten 30-minute intervals (black curves), showing ({\bf left column, top to bottom})  the normalized fluctuating total density $\delta \rho$, fluctuating velocity components $\delta u_{z,y,x}$, and ({\bf right column, top to bottom}) pressure $\delta p$, and fluctuating magnetic field components $\delta B_{z,y,x}$ respectively.  All fluctuating quantities are normalized to the appropriate mean values as labeled. } \label{fig:8}
\end{figure}

A final test of the accuracy of the decomposition is illustrated in Figure \ref{fig:8}. Here, each of the mode-decomposed modal contributions to the normalized density, pressure, velocity, and magnetic field fluctuations are reassembled to determine the total normalized density, pressure, velocity, and magnetic field fluctuations. The recomposed total plasma and magnetic field variables are then compared to the corresponding original measured fluctuations. The measured fluctuations are illustrated in orange and labeled with the superscript ``$m$'' and the reassembled plasma and magnetic field fluctuations are shown by the black curve. The reconstruction of the normalized data closely reproduces the observed data, being indistinguishable for much of the data set, giving us confidence in the efficacy of the mode-decomposition method and assumptions. 

\section{Discussion and conclusions} \label{sec:4}

The characterization of fluctuations in the solar wind provides considerable insight into the nature of turbulence in a magnetofluid. By means of a novel extension of mode-decomposition \citep{Zank_etal_2023}, we can identify and characterize the properties of entropy, magnetic islands, forward and backward Alfv\'en waves, including weakly or non-propagating Alfv\'en vortices, and forward and backward fast and slow magnetosonic modes. The analysis presented here is not a theory of turbulence but merely characterized the fluctuations within a short time interval, the ``decoherence time,'' during which the fluctuations in a plasma parcel behave essentially linearly while in the parcel and do not experience nonlinear interactions. This snapshot of the fluctuation  characteristics of the plasma parcel has the added advantage of allowing us to exploit the dispersion relation of each mode to relate the frequency to the wavenumber and hence construct wavenumber spectra or PSDs for each mode from frequency spectra obtained from spacecraft observations. Our mode-decomposition approach \citep{Zank_etal_2023} therefore provides surprisingly detailed insights into the fundamental building blocks of a compressible turbulent magnetofluid. 

In this work, we investigate fluctuations in the first sub-Alfv\'enic solar wind flow observed by PSP \citep{Kasper_etal_2021, Zank_etal_2022, Zhao_etal_2022a, Bandyopadhyay_etal_2022} for 5 hours on 2021 April 28. Our results, related discussion, and conclusions are listed below. 

\begin{enumerate}
	
	\item The mode-decomposition has identified density fluctuations associated with entropy modes in the sub-Alfv\'enic flow. In addition, we  identified density fluctuations associated with forward and backward propagating fast and slow magnetosonic modes. 
	The density fluctuations in this sub-Alfv\'enic flow are comprised primarily and almost equally of entropy and backward propagating slow magnetosonic (ms) modes. In a small plasma beta environment, the fast ms mode is essentially a magnetic fluctuation whereas the slow ms mode, with a phase velocity given approximately by $V_p^{s\pm} \simeq \pm V_{A0} \sqrt{(\gamma/2) \beta_p} \cos \theta^{s\pm} = \left( \gamma P_0/\rho_0 \right)^{1/2} \cos \theta^{s\pm} = C_{S0}  \cos \theta^{s\pm}$ (where $C_{S0} = \sqrt{\gamma P_0/\rho_0}$ is the plasma sound speed) is almost a pure sound wave. Hence, the fast ms mode scarcely contributes to the fluctuating density, which is due primarily to the slow ms and entropy modes. 
	Both the slow ms and entropy density fluctuations possess a $k^{-5/3}$ wavenumber spectrum, the latter of which is indicative of fully developed MHD turbulence and a consequence of turbulent advection by quasi-2D velocity fluctuations. Such a spectrum appears to be consistent with the evolution of density fluctuations measured remotely by the Solar Orbiter instrument Metis that show that the density spectrum evolves towards a Kolmogorov-like spectrum by about 3 R${}_{\odot}$ \citep{Telloni_etal_2023a}, a region that is certainly sub-Alfv\'enic. Such a spectrum indicates that the MHD turbulence is fully developed by about 3 R${}_{\odot}$. Both dominant density spectra show some evidence of weak or slight flattening at larger wavenumbers. 
	
	Density fluctuations in the solar corona are particularly important in the context of both solar radio bursts and the angular broadening and scintillation of galactic and extra-galactic compact radio sources. Since propagating radio waves are strongly affected by scattering, this affects the observed time characteristics, sizes, and positions of solar radio bursts and compact radio sources (e.g., \cite{Kontar_etal_2023} and references therein). In particular, radio observations suggest that the density fluctuations are anisotropic in that parallel wavenumbers are smaller than perpendicular wavenumbers (e.g., \cite{Coles_Harmon_1989, Armstrong_etal_1990}), with possibly a more pronounced  anisotropy at smaller scales. For the densities associated with entropy fluctuations, we find a median value of $\theta^e = 166.7^{\circ}$ in the sub-Alfv\'enic interval, which yields a median anisotropy of  $|k_{\parallel} / k_{\perp}| = 4.2$ (since $\theta^e \in [143.3^{\circ}, 177.8^{\circ}]$, we have $|k_{\parallel} / k_{\perp}| \in [1.3, 26]$). Thus, $k_{\parallel} \gg k_{\perp}$ for advected entropy-density fluctuations, which reflects the close alignment of the mean flow with the mean magnetic field. Hence, the wavenumber anisotropy is in the opposite sense of what is expected from or required for radio wave scattering in the corona. Consider now the other dominant density component, the backward propagating slow ms mode, which has a median value of $\theta^{s-} = 124.8^{\circ}$ and $\theta^{s-}$ ranges from $[123.31^{\circ}, 126.86^{\circ}]$. This yields a median wavenumber anisotropy of  $|k_{\parallel} / k_{\perp}| = 0.7$ with a range of $|k_{\parallel} / k_{\perp}| \in [0.65, 0.75]$. Thus, the wavenumber anisotropy of the backward propagating slow ms modes is in the sense of $k_{\perp} > k_{\parallel}$. This range of values is larger than identified by \cite{Kontar_etal_2023} who needed $k_{\parallel}/k_{\perp} = 0.25 - 0.4$ to account for the shortest solar radio burst decay times observed. Since the strongest contribution to the scattering of radio waves is at the ion inertial scale, perhaps about 10 km, this an order of magnitude smaller than the smallest length scale in the mode-decomposition analysis. Thus, it is entirely possible that the anisotropy around the ion inertial scales may be larger. 
	Finally, the variance anisotropy of the density associated with the entropy, backward slow, and backward fast modes is 4.16, 0.68, and 0.68 respectively. The wavenumber and variance  anisotropy values hold over the wavenumber range $2 \times 10^{-6} \leq k \leq 10^{-2}$ km${}^{-1}$.
	
	\item The mode-decomposition allowed us to identify magnetic islands, forward and backward propagating Alfv\'en waves, including Alfv\'en vortices, and forward and backward propagating fast and slow magnetosonic modes. Both the Alfv\'en and fast ms waves are essentially uni-directionally propagating, with the forward propagating mode dominant in both cases. However, the backward slow ms is the dominant slow mode. The dominant mode for all fluctuations identified is magnetic islands. The dominance of the magnetic islands is most apparent in the spectral plots and we find that the variance anisotropy is $\langle {\delta B^i}^2 \rangle / \langle {\delta B^A}^2 \rangle = 4.1$, which is consistent with the predictions of NI MHD in the small and $O(1)$ plasma beta regimes \citep{zank_matthaeus_1993_incompress, Zank_etal_2020} and corresponds to the $80\%:20\%$ ratio for 2D:slab fluctuations derived by \cite{Bieber_etal_1996JGR, Sauer_Bieber_1999} from observations at 1 au. 
	
	Finally, the spectral amplitudes  per logarithmic wavenumber for incompressible (transverse) magnetic fluctuations associated with magnetic islands and Alfv\'en waves are 10 (forward fast ms mode) and 100 (backward slow ms mode) times greater than the corresponding spectral amplitudes for compressible fluctuations. Thus, in terms of magnetic field fluctuations, the sub-Alfv\'enic flow is essentially incompressible with longitudinal fluctuations being a minority component. However, despite the forward fast magnetosonic mode having a spectral amplitude in the fluctuating magnetic variance nearly an order of magnitude greater than that of the backward propagating slow ms mode, the ordering of the spectral amplitudes for the density variance is reversed, and dominated by the backward slow ms mode. 
	
	The incompressible magnetic island spectrum is given by a Kolmogorov-like $k_{\perp}^{-1.6}$ spectrum, consistent with the expectations of the spectral theory for NI MHD \citep{Zank_etal_2020}. The idealized 2D spectrum obviously does not possess $k_{\parallel}$ wavenumbers. The spectral forms for the uni-directionally propagating (forward) Alfv\'en modes are $k_{\parallel}^{-1.6}$ and $k_{\perp}^{-1.5}$, with the latter spectrum exhibiting a somewhat convex profile. The formation and power law shape of the uni-directionally propagating Alfv\'en wave spectrum cannot be a consequence of oppositely propagating Alfv\'en waves, implying an apparent absence of nonlinear interactions to initiate the cascade that can form an Irshnikov-Kraichnan $k^{-3/2}$ spectrum. As shown in \cite{Zank_etal_2020}, frequency sweeping (referred to as scattering in \cite{Zank_etal_2020}) of a uni-directional Alfv\'en wave distribution by the dominant 2D fluctuations can result in a power law distribution.  
	
	For the compressible ms modes, the energy in the longitudinal magnetic component is slightly larger than that of the transverse components. Both the longitudinal and transverse variances appear to have similar magnetic power law spectra with spectral indices flatter than either $-5/3$ or even $-3/2$. 
	
	Besides being essentially uni-directional, the Alfv\'en and fast ms modes both propagate at essentially the Alfv\'en speed $V_{A0}$ (the correction to the fast ms phase speed being $\sim \beta_p$ since $V_p^{f\pm} \simeq V_{A0} \left[ 1 + (\gamma /4) \beta_p \sin^2 \theta^{f\pm} \right]$).  
	Furthermore, although the longitudinal component of the fast ms mode magnetic field fluctuations is only modestly larger per logarithmic wavenumber than the transverse component, the corresponding transverse velocity components are more than two orders of magnitude larger in spectral amplitude per logarithmic wavenumber than the longitudinal velocity component. Consequently, the transverse magnetic and velocity fluctuating components are essentially anti-correlated (see Figures \ref{fig:3} and \ref{fig:6}). A simple analysis of the velocity and magnetic field data would therefore incorrectly ``identify'' the fast ms mode as being an outwardly propagating Alfv\'en wave. This is of particular interest in the context of characterizing the underlying nature of switchbacks \citep{Kasper_etal_2019}, which are typically taken to be Alfv\'enic based on the correlation of transverse magnetic and velocity fluctuations \citep{Kasper_etal_2019, Fisk_Kasper_2020, Tenerani_etal_2020, McManus_etal_2020, Dudok_Wit_etal_2020}. However, switchbacks possess a longitudinal magnetic field and velocity component, often correlated, which is inconsistent with an Alfv\'en wave but is consistent with a fast ms mode, leading \cite{Zank_etal_2020b} to argue that switchbacks were rather fast ms mode structures propagating in a small plasma beta solar wind. As shown here,  fast ms modes in a low plasma beta environment are almost indistinguishable from an Alfv\'en mode, both propagating at $V_{A0}$, both having anti-correlated transverse magnetic and velocity components, but only the fast mode possesses a longitudinal magnetic and velocity field component. 
	
	\item The mode-decomposition enables a close examination of the velocity fluctuations, which reveals surprisingly complicated characteristics. For example, the forward fast ms mode is dominated by the transverse velocity fluctuations, unlike the magnetic field fluctuations for which the  longitudinal component is slightly larger in spectral amplitude per logarithmic wavenumber. By contrast, the backward propagating slow ms mode is dominated by the longitudinal velocity fluctuations and not the transverse components, consistent with the magnetic field fluctuations.  As discussed above, this is simply a reflection of the small plasma beta, which renders the slow mode as essentially a sound wave modified only weakly by the magnetic field and the sound wave as essentially a magnetic wave very similar to the Alfv\'en mode. The fast and slow ms velocity PSDs are not as flat as those for the magnetic field fluctuations. Unlike the magnetic field fluctuations, the incompressible velocity fluctuations are not dominant and the kinetic energy density of the backward slow ms mode is comparable to that of the forward Alfv\'en mode. This is likely to make it challenging to interpret velocity or kinetic energy spectra compared to the more clearly distinguishable magnetic field PSDs that are dominated by the incompressible magnetic field component. However, the mode-decomposition analysis allows one to separate the transverse velocity components from the fast and slow ms components and so construct the kinetic energy PSD of the transverse velocity components (i.e., the Alfv\'en velocity fluctuations and the fast and slow ms mode transverse velocity fluctuations) exclusively. In so doing, we found that, although differing in spectral amplitude by a factor of $\sim 2$, the fluctuating transverse kinetic energy and the magnetic field (expressed in Alfv\'en units) PSDs were almost identical. This is a restatement that the fast ms mode is almost Alfv\'enic in that the energy density in transverse magnetic and transverse velocity fluctuations is equal, indicating again that short of applying a mode-decomposition analysis, it is difficult to distinguish between Alfv\'en and fast ms modes in a low plasma beta region.
	In summary, the mode-decomposition analysis suggests caution in interpreting kinetic energy spectra.
	
	\item We have compared the frequency spectra for transverse magnetic field fluctuations derived from a Fourier transform of the original data with that derived from the mode-decomposition analysis for both the 30-minute subintervals and the full 5-hour interval. The agreement in spectral amplitude and features for the individual 30-minute subintervals is excellent. The construction of a 5-hour mode-decomposed frequency PSD from the 10 30-minute subintervals is effectively a form of ``ensemble averaging'' of 10 realizations since the mean or background plasma parameters are very similar for each subinterval. The agreement between the Fourier-derived and the mode-decomposition-derived PSDs for the 5-hour interval is good with the basic features matching well but there is a small difference in spectral amplitude per logarithmic wavenumber and the high-frequency 
	part of the spectra differ in that one flattens and the other steepens slightly. Nonetheless, these differences are not very significant. This indicates that our choice of the ``decoherence time'' $\Delta t$ based on the confinement time of fluctuations within a subinterval and on the nonlinear and Alfv\'en timescales was reasonable. 
	A comparison of the time series data based on a reconstruction from the mode-decomposition shows excellent agreement with the original plasma data. Both the time series and spectral comparisons give us confidence that mode-decomposition provides an accurate snapshot and classification of the fluctuations comprising a plasma parcel of the sub-Alfv\'enic solar wind. 
	
	A final point that we did not address in great detail but which nonetheless warrants some consideration is in our use of 30-min averaged mean magnetic and plasma variables rather than  localized values. Although ``semi-localized'' compared to using 5-hour mean values, it is nonetheless a global mean magnetic field and it is well-known that using global- or local-mean
	field coordinates can have a significant effect in analyzing the variance anisotropy \citep[e.g.,][]{Oughton_etal_2015}.  \cite{Chaston_etal_2020}, for example, used a local mean value coordinate system at each frequency in the spacecraft frame and time in their mode-decomposition analysis, i.e., they determined a local mean magnetic field ${\bf B}_0$ at each wave scale of interest. This of course has important implications for the determination of the wavenumber angles that we calculate in our mode-decomposition analysis. While it would be interesting to explore the use of local mean values, especially the mean magnetic field from which the coordinate system is drawn, there is the danger that one may be violating the basic nature of the mode-decomposition method since linearization is done on the basis of a mean background. A more detailed analysis would be of interest in future work.
	
\end{enumerate}

\acknowledgments
\begin{acknowledgements}
\section*{ACKNOWLEDGMENTS}
GPZ, LZ, LA, MN, PB, PB, and XZ acknowledge the partial support of a NASA Parker Solar Probe contract SV4-84017 and an NSF EPSCoR RII-Track-1 Cooperative Agreement OIA - 2148653. A.P. acknowledges the partial support of the Czech Grant Agency under contract (23-06401S).
\end{acknowledgements}

\section*{APPENDIX}
Several typos are present in some of the equations listed in \cite{Zank_etal_2023}. Here we provide the corrected equations needed for the mode-decomposition algorithm. A  corrected version of  \cite{Zank_etal_2023}, including one of the derivations, is available by request from any of the authors listed here. 

The mode-decomposition requires the inversion of the linear equation containing the amplitude matrix ${\bf A}$, adopting the geometry of Figure 1 in \cite{Zank_etal_2023}, given by 
\begin{eqnarray}
{\bf A} {\bf x} = {\bf b}, \quad \mbox{where} \quad {\bf A} = (a_{ij}), {\bf x} = (x_j), {\bf b} = (b_i), \quad i,j = 1 \cdots 8; \nonumber \\
{\bf x} = \left( \frac{\delta \rho^e }{\rho_0}, \frac{\delta \hat{p}^{f+} }{\rho_0 a_0^2}, \frac{\delta \hat{p}^{f-} }{\rho_0 a_0^2}, \frac{\delta \hat{p}^{s+} }{\rho_0 a_0^2}, \frac{\delta \hat{p}^{s-} }{\rho_0 a_0^2},  \frac{\delta u^{A+} }{U_0}, \frac{\delta u^{A-} }{U_0}, \frac{\delta B^i}{B_0}  \right)^t.  \label{eq:A1}  
\end{eqnarray}
The amplitude matrix contains the amplitudes and phases of the entropy, fast, and slow magnetosonic, Alfv\'enic, and magnetic island modes, and the vector ${\bf b}$ is comprised of measured plasma and magnetic field values and is given below.  The elements $(a_{ij})$ are derived from the MHD conservation laws and listed below -- the reader is referred to  Appendix B in \cite{Zank_etal_2023} for the definitions of the various terms.

\begin{landscape}
\begin{equation}
\begin{array}{llllllll}
a_{11} = 1 & a_{12} = a_{12}(f+) & a_{13} = a_{12}(f-) & a_{14} = a_{12}(s+) & a_{15} = a_{12}(s-) & a_{16} = a_{16}(A^+) & a_{17} = a_{16}(A^-)& a_{18} = 0 \\
a_{21} = \sin \psi & a_{22} = a_{22}(f+) & a_{23} = a_{22}(f-) & a_{24} = a_{22}(s+) & a_{25} = a_{22}(s-) & a_{26}  &  a_{27} & a_{28} = \frac{\cos \psi }{M_{A0}^2} \beta^i \\
a_{31} = 0 & a_{32} = a_{32} (f+) & a_{33} = a_{32} (f-) & a_{34} = a_{32} (s+) & a_{35} = a_{32} (s-) & a_{36} & a_{37} & a_{38} = - \frac{\cos \psi}{M_{A0}^2 } \alpha^i \\
a_{41} = \cos \psi & a_{42} = a_{42} (f+) & a_{43} = a_{42} (f-) & a_{44} = a_{42} (s+) & a_{45} = a_{42} (s-) & a_{46} & a_{47} & a_{48} = \frac{\sin \psi}{M_{A0}^2}  \beta^i \\ 
a_{51} = \frac{1}{2} & a_{52} = a_{52} (f+) & a_{53} = a_{52} (f-) & a_{54} = a_{52} (s+) & a_{55} = a_{52} (s-) & a_{56} & a_{57} & a_{58} = 2\frac{\sin \psi \cos \psi}{M_{A0}^2} \beta^i \\
a_{61} = 0 & a_{62} = a_{62} (f+) & a_{63} = a_{62} (f-) & a_{64} = a_{62} (s+) & a_{65} = a_{62} (s-) & a_{66} & a_{67} & a_{68} = \cos^2 \psi \beta^i \\ 
a_{71} = 0 & a_{72} = a_{72} (f+) & a_{73} = a_{72} (f-) & a_{74} = a_{72} (s+) & a_{75} = a_{72} (s-) & a_{76} & a_{77} & a_{78} = -\alpha^i \\
a_{81} = 0 & a_{82} = a_{82} (f+) & a_{83} = a_{82} (f-) & a_{84} = a_{82} (s+) & a_{85} = a_{82} (s-) & a_{86} = M_{A0} \beta^{A+} & a_{87} = -M_{A0} \beta^{A-} & a_{88} = -\beta^i 
\end{array} \label{eq:A2} 
\end{equation}
\end{landscape}
where 
\begin{eqnarray*}
	a_{12} &=& \left( 1 + \frac{1}{M_0} \frac{V_{f+} }{a_0} \frac{V_{f}^2 \cos \phi^{f+} \sin \theta^{f+} }{V_f^2 - V_{A0}^2 \cos^2 \theta^{f+} } \sin \psi + \frac{1}{M_0} \frac{a_0}{V_{f+} } \cos \theta^{f+} \cos \psi \right) = a_{12}(f+) ; \\
	a_{16} & = & -\beta^{A+} \sin \psi = a_{16}(A^+) ; \\ 
	a_{22} &=& \sin \psi \left( 1 + \frac{1}{M_0^2} \right) + \frac{1 + \sin^2 \psi}{M_0} \frac{V_{f+} }{a_0} \frac{V_f^2 \cos \phi^{f+} \sin \theta^{f+} }{V_f^2 - V_{A0}^2 \cos^2 \theta^{f+} } + \frac{\sin \psi \cos \psi}{M_0} \frac{a_0}{V_{f+} } \cos \theta^{f+} \\
	&\mbox{}& \mbox{} + \frac{\sin \psi}{M_{A0}^2 } \frac{V_f^2 - a_0^2 \cos^2 \theta^{f+} }{v_f^2} + \frac{\cos \psi}{M_{A0}^2 } \frac{V_f^2 \cos \phi^{f+} \sin \theta^{f+} \cos \theta^{f+} }{V_f^2 - V_{A0}^2 \cos^2 \theta^{f+} }; \\
	a_{26} &=& -\left( \frac{ \cos \psi}{M_{A0} } + (1 + \sin^2 \psi) \right) \beta^{A+} ; \quad 
	a_{27} = \left( \frac{ \cos \psi}{M_{A0} } - (1 + \sin^2 \psi) \right) \beta^{A-} ; \\
	a_{32} &=& \frac{V_f^2 \sin \phi^{f+} \sin \theta^{f+} }{V_f^2 - V_{A0}^2 \cos^2 \theta^{f+} } \left( \frac{\cos \psi}{M_{A0}^2 } \cos \theta^{f+} + \frac{1}{M_{f+} (\theta^{f+})} \right); \quad 
	a_{36} = \left( 1 + \frac{\cos \psi}{M_{A0} } \right) \alpha^{A+}; \\
	a_{37} & = & \left( 1 - \frac{\cos \psi}{M_{A0} } \right) \alpha^{A-}; \\ 
	a_{42} &=& \cos \psi \left( 1 + \frac{1}{M_0^2} \right) + \frac{\cos \psi \sin \psi}{M_0} \frac{V_{f+} }{a_0} \frac{V_f^2 \cos \phi^{f+} \sin \theta^{f+} }{V_f^2 - V_{A0}^2 \cos^2 \theta^{f+} } + (1 + \cos^2 \psi ) \frac{a_0/V_{f+}}{M_0} \cos \theta^{f+} \\
	&\mbox{}& \mbox{} + \frac{\sin \psi}{M_{A0}^2 } \frac{V_f^2 \cos \phi^{f+} \sin \theta^{f+} \cos \theta^{f+} }{V_f^2 - V_{A0}^2 \cos^2 \theta^{f+} } - \frac{\cos \psi}{M_{A0}^2 } \frac{V_f^2 - a_0^2 \cos^2 \theta^{f+} }{V_f^2} ; \\ 
	a_{46} &=& -\sin \psi \left( \cos \psi + \frac{1}{M_{A0} } \right) \beta^{A+}; \quad 
	a_{47} = -\sin \psi \left( \cos \psi - \frac{1}{M_{A0} } \right) \beta^{A-};  \\ 
	a_{52} &=& \frac{1}{2} + \frac{\gamma}{\gamma - 1} \frac{1}{M_0^2} + \left( \frac{{\cal E}_T}{U_0^2} + 1 \right) \frac{\sin \psi}{M_0} \frac{V_{f+}}{a_0} \frac{V_f^2 \cos \phi^{f+} \sin \theta^{f+} }{V_f^2 - V_{A0}^2 \cos^2 \theta^{f+} } + \left( \frac{{\cal E}_0}{U_0^2} + 1 \right) \frac{\cos \psi}{M_0} \frac{a_0}{V_{f+}} \cos \theta^{f+} \\
	&\mbox{}& \mbox{} + 2\frac{\sin \psi \cos \psi}{M_{A0}^2 } \frac{V_f^2 \cos \phi^{f+} \sin \theta^{f+} \cos \theta^{f+} }{V_f^2 - V_{A0}^2 \cos^2 \theta^{f+} } +2 \frac{\sin^2 \psi}{M_{A0}^2} \frac{V_f^2 - a_0^2 \cos^2 \theta^{f+} }{V_f^2}; \\ 
	a_{56} &=& -\left( \left( \frac{{\cal E}_T}{U_0^2} + 1 \right) + 2\frac{\cos \psi}{M_{A0}} \right) \sin \psi \beta^{A+}; \quad 
	a_{57} = -\left( \left( \frac{{\cal E}_T}{U_0^2} + 1 \right) - 2\frac{\cos \psi}{M_{A0}} \right) \sin \psi \beta^{A-}; \\ 
	a_{62} &=& \cos \psi \left( \frac{V_f^2 \cos \phi^{f+} \sin \theta^{f+}}{V_f^2 - V_{A0}^2 \cos^2 \theta^{f+}} \left( \frac{V_{f+}/a_0}{M_0} + \cos \psi \cos \theta^{f+} \right) + \sin \psi \frac{V_f^2 - a_0^2 \cos^2 \theta^{f+}}{V_f^2} \right); \\
	a_{66} &=& -\cos \psi \left( 1 + \cos \psi M_{A0} \right) \beta^{A+}; \quad 
	a_{67} = -\cos \psi \left( 1 - \cos \psi M_{A0} \right) \beta^{A-}; \\
	a_{72} &=& \frac{V_f^2 \sin \phi^{f+} \sin \theta^{f+}}{V_f^2 - V_{A0}^2 \cos^2 \theta^{f+}} \left( \frac{\cos \psi}{M_0} \frac{V_{f+}}{a_0} + \cos \theta^{f+} \right); \quad a_{76} = \left( \cos \psi + M_{A0} \right) \alpha^{A+};\\
	a_{77} & = & \left( \cos \psi - M_{A0} \right) \alpha^{A-}; \\
	a_{82} &=& -\frac{V_f^2 \cos \phi^{f+} \sin \theta^{f+} \cos \theta^{f+} }{V_f^2 - V_{A0}^2 \cos^2 \theta^{f+}} 
\end{eqnarray*} 
and ${\cal E}_0 \equiv (1/2) U_0^2 + a_0^2/(\gamma - 1)$ and ${\cal E}_T \equiv {\cal E}_0 + V_{A0}^2$.

The source vector $(b_i)$, $i = 1 \cdots 8$, is determined from the measured plasma and magnetic field variables, denoted by the subscript $m$.
\begin{eqnarray*}
	b_1 &=& \frac{\delta \hat{\rho}_m}{\rho_0} + \sin \psi \frac{\delta \hat{u}_{mx}}{U_0} + \cos \psi \frac{\delta \hat{u}_{mz}}{U_0}; \\ 
	b_2 &=& \sin \psi \frac{\delta \hat{\rho}_m}{\rho_0} + \left( 1 + \sin^2 \psi \right) \frac{\delta \hat{u}_{mx}}{U_0} + \sin \psi \cos \psi \frac{\delta \hat{u}_{mz}}{U_0} + \frac{\sin \psi}{M_0^2} \frac{\delta \hat{p}_m}{\rho_0 a_0^2} \\
	 &+& \frac{1}{M_{A0}^2} \left( -\cos \psi \frac{\delta \hat{B}_{mx}}{B_0} + \sin \psi \frac{\delta \hat{B}_{mz}}{B_0} \right); \\
	b_3 &=& \frac{\delta \hat{u}_{my}}{U_0} - \frac{\cos \psi}{M_{A0}^2} \frac{\delta \hat{B}_{my}}{B_0}; \\
	b_4 &=& \cos \psi \frac{\delta \hat{\rho}_m}{\rho_0} + \cos \psi \sin \psi \frac{\delta \hat{u}_{mx}}{U_0} + \left( 1 + \cos^2 \psi \right) \frac{\delta \hat{u}_{mz}}{U_0} + \frac{\cos \psi}{M_0^2} \frac{\delta \hat{p}_m}{\rho_0 a_0^2}\\
	 & - & \frac{1}{M_{A0}^2} \left( \sin \psi \frac{\delta \hat{B}_{mx}}{B_0} + \cos \psi \frac{\delta \hat{B}_{mz}}{B_0} \right); \\
	b_5 &=& \frac{1}{2} \frac{\delta \hat{\rho}_m}{\rho_0} + \left( \frac{{\cal E}_T}{U_0^2} + 1 \right) \sin \psi \frac{\delta \hat{u}_{mx}}{U_0} + \left( \frac{{\cal E}_0}{U_0^2} + 1 \right) \cos \psi \frac{\delta \hat{u}_{mz}}{U_0} + \frac{\gamma}{\gamma - 1} \frac{1}{M_0^2}  \frac{\delta \hat{p}_m}{\rho_0 a_0^2} \\
	& - & 2\frac{\sin \psi \cos \psi}{M_{A0}^2} \frac{\delta \hat{B}_{mx}}{B_0} + 2\frac{\sin^2 \psi }{M_{A0}^2} \frac{\delta \hat{B}_{mz}}{B_0}; \\
	b_6 &=& \cos \psi \frac{\delta \hat{u}_{mx}}{U_0} - \cos^2 \psi  \frac{\delta \hat{B}_{mx}}{B_0} + \sin \psi \cos \psi  \frac{\delta \hat{B}_{mz}}{B_0}; \\
	b_7 &=& \cos \psi \frac{\delta \hat{u}_{my}}{U_0} - \frac{\delta \hat{B}_{my}}{B_0};  \\
	b_8 &=& \frac{\delta \hat{B}_{mx}}{B_0}. 
\end{eqnarray*}

The angles that are needed for the $8 \times 8$ amplitude matrix ${\bf A}$ are obtained sequentially from the following system of equations, 
\begin{eqnarray*}
	&\bullet& \quad \cos^2 \theta^{f+} = \frac{1}{M_0^2 M_{A0}^2} \frac{\left( M_0^2 + M_{A0}^2 \right) \left( \omega_m^{\prime}/(U_0 k_{mz} )\right)^2 - 1}{\left( \omega_m^{\prime}/(U_0 k_{mz} )\right)^4 }; \\
	&\bullet&\quad  \cos \phi^{f+} = \frac{1}{ \left( \omega_m^{\prime}/(U_0 k_{mx}) \right) M_f(\theta^{f+}) \sin \theta^{f+} } \\
	&\bullet& \quad \theta^{f-} =  \pi - \theta^{f+}; \\
	&\bullet& \quad \phi^{f-} = \pi -  \phi^{f+}; \\
	&\bullet& \quad \cos^2 \theta^{s+} = \frac{M_f^2(\theta^{f+}) }{M_0^2 M_{A0}^2} \cos^2 \theta^{f+} \left( M_0^2 + M_{A0}^2 - M_f^2 (\theta^{f+}) \cos^2 \theta^{f+} \right); \\
	&\bullet& \quad \cos \phi^{s+} = \frac{M_f(\theta^{f+})}{M_s(\theta^{s+})} \frac{\cos \phi^{f+} \sin \theta^{f+}}{\sin \theta^{s+}}; \\
	&\bullet& \quad \theta^{s-} = \pi - \theta^{s+}; \\
	&\bullet& \quad \phi^{s-} =  \pi - \phi^{s+}; \\
	&\bullet& \quad \phi^{e} = 0; \\
	&\bullet& \quad \tan \theta^e = \frac{\sin \psi + M_f(\theta^{f+}) \cos \phi^{f+} \sin \theta^{f+}}{\cos \psi + M_f(\theta^{f+}) \cos \theta^{f+}}. 
\end{eqnarray*}
The angles $\phi^{A\pm}$, $\theta^{A\pm}$, and $\phi^i$ associated with the Alfv\'enic and magnetic island modes are solved iteratively from 
\begin{eqnarray*}
	&\bullet& \quad \frac{\delta \hat{u}^{A+} }{U_0} \sin \phi^{A+} +  \frac{\delta \hat{u}^{A-} }{U_0} \sin \phi^{A-} = - \frac{\delta \hat{u}_{mx} }{U_0} + \frac{1}{M_f(\theta^{f+})} \left[ \frac{\delta \hat{p}^{f+} }{\rho_0 a_0^2} - \frac{\delta \hat{p}^{f-} }{\rho_0 a_0^2} + \frac{M_f^2 (\theta^{f+}) }{M_s^2 (\theta^{s+})} \left( \frac{\delta \hat{p}^{s+} }{\rho_0 a_0^2} - \frac{\delta \hat{p}^{s-} }{\rho_0 a_0^2} \right) \right]  \\
	&\mbox{}& \mbox{} \times \frac{M_{A0}^2 \cos \phi^{f+} \sin \theta^{f+} }{M_{A0}^2 - M_f^2 (\theta^{f+}) \cos^2 \theta^{f+} }; \\
	&\bullet& \quad \frac{\delta \hat{u}^{A+} }{U_0} \cos \phi^{A+} +  \frac{\delta \hat{u}^{A-} }{U_0} \cos \phi^{A-} = \frac{\delta \hat{u}_{my} }{U_0} - \left( \frac{\delta \hat{p}^{f+} }{\rho_0 a_0^2} + \frac{\delta \hat{p}^{f-} }{\rho_0 a_0^2} \right)  \frac{1}{M_f(\theta^{f+})}  \frac{M_{A0}^2 \sin \phi^{f+} \sin \theta^{f+} }{M_{A0}^2 - M_f^2 (\theta^{f+}) \cos^2 \theta^{f+} } \nonumber \\ 
	&\mbox{}& \mbox{} - \left( \frac{\delta \hat{p}^{s+} }{\rho_0 a_0^2} + \frac{\delta \hat{p}^{s-} }{\rho_0 a_0^2} \right)   \frac{1}{M_s(\theta^{s+})}  \frac{M_{A0}^2 \sin \phi^{s+} \sin \theta^{s+} }{M_{A0}^2 - M_s^2 (\theta^{s+}) \cos^2 \theta^{s+} }; \\
	&\bullet& \quad \tan \theta^{A\pm} = \left( \cos \phi^{A\pm} \right)^{-1} \frac{ M_{A0} \cos \psi \pm 1}{M_{A0} \cos \psi} \frac{\sin \psi + M_f(\theta^{f+}) \cos \phi^{f+} \sin \theta^{f+} }{\cos \psi + M_f(\theta^{f+}) \cos \theta^{f+} }; \\
	&\bullet& \quad  \frac{\delta \hat{B}^i}{B_0} \cos \phi^i = \frac{\delta \hat{B}_{my} }{B_0} + \left( \frac{\delta \hat{p}^{f+} }{\rho_0 a_0^2} - \frac{\delta \hat{p}^{f-} }{\rho_0 a_0^2} \right) \frac{M_{A0}^2 \sin \phi^{f+} \sin \theta^{f+} \cos \theta^{f+} }{M_{A0}^2 - M_f^2 (\theta^{f+}) \cos^2 \theta^{f+} } \\
	&\mbox{}& \mbox{} +  \left( \frac{\delta \hat{p}^{s+} }{\rho_0 a_0^2} - \frac{\delta \hat{p}^{s-} }{\rho_0 a_0^2} \right) \frac{M_{A0}^2 \sin \phi^{s+} \sin \theta^{s+} \cos \theta^{s+} }{M_{A0}^2 - M_s^2 (\theta^{s+}) \cos^2 \theta^{s+} }  + M_{A0}  \left( \frac{\delta \hat{u}^{A+} }{U_0} \cos \phi^{A+} - \frac{\delta \hat{u}^{A-} }{U_0} \cos \phi^{A-} \right) 
\end{eqnarray*}

Finally, for convenience, we list the conversions that map frequency to wavenumber. The equation numbers below refer to the equations in \cite{Zank_etal_2023}. \\

\noindent
\underline{Entropy modes}: For each $\omega_m^{\prime}$, we map to the entropy wavenumber ${\bf k}^e$ using equations  (95), (96), $k_y^e = 0$, and (97), (98) in \cite{Zank_etal_2023}. \\

\noindent
\underline{Magnetic islands}: For each $\omega_m^{\prime}$, $k_x^i$ is given by (109), $\phi^i$ by (110) \citep{Zank_etal_2023}, and $k^i$ is determined from $k_x^i = k^i \cos \phi^i$ {\bf (since $\theta^i = \pi/2$)}. \\

\noindent
\underline{Fast/slow magnetosonic modes}: For each of these four cases, we have $k^{fs\pm} = \omega_m^{\prime}/V_{fs} (\theta^{fs\pm})$ and hence $k_{\perp}^{fs\pm} = k^{fs\pm} \sin \theta^{fs\pm}$ and $k_{\parallel}^{fs\pm} = k^{fs\pm} \cos \theta^{fs\pm}$. \\

\noindent
\underline{Alfv\'en modes}: For each $\omega_m^{\prime}$, equations (106) and (107) provide $\phi^{A\pm}$ and (108) \citep{Zank_etal_2023} gives $\theta^{A\pm}$. Since $k_x^{A+} = k_x^{A-}$, we have 
\begin{equation}
k^{A\pm} = \left( \cos \phi^{A\pm} \sin \theta^{A\pm} \right)^{-1} \frac{\omega_m^{\prime} }{U_0} \left( \sin \psi + M_f(\theta^{f+}) \cos \phi^{f+} \sin \theta^{f+} \right), \label{eq:120}
\end{equation}
giving $k_{\perp}^{A\pm} = k^{A\pm} \sin \theta^{A\pm}$ and $k_{\parallel}^{A\pm} = k^{A\pm} \cos \theta^{A\pm}$.

\providecommand{\noopsort}[1]{}\providecommand{\singleletter}[1]{#1}%


\end{document}